\documentclass[aps, prd, 11pt, twocolumn, nofootinbib, superscriptaddress]{revtex4-2}
\pdfoutput=1
\usepackage{microtype}
\usepackage{float}
\usepackage{amsfonts}
\usepackage{psfrag}
\usepackage{graphicx}
\usepackage{grffile}
\usepackage{cancel}
\usepackage{amsmath}
\usepackage{natbib}
\usepackage{xr}
\usepackage{hyperref}
\usepackage{verbatim}
\usepackage{epstopdf}
\usepackage{subcaption}
\usepackage{caption}
\usepackage{multirow}
\usepackage{float}
\usepackage{commath}
\usepackage{empheq}
\usepackage{braket}
\usepackage{array}
\usepackage{booktabs,dcolumn,caption}
\hypersetup{
    bookmarks=true,         
    unicode=false,          
    pdftoolbar=true,        
    pdfmenubar=true,        
    pdffitwindow=false,     
    pdfstartview={FitH},    
    pdfauthor={Nathan Rutherford},     
    colorlinks=true,       
    linkcolor=blue,          
    citecolor=blue,        
}
\newcolumntype{d}[1]{D{.}{.}{#1}}

\RequirePackage[normalem]{ulem} 
\RequirePackage{color}\definecolor{RED}{rgb}{1,0,0}\definecolor{BLUE}{rgb}{0,0,1} 
\providecommand{\DIFaddbegin}{} 
\providecommand{\DIFaddend}{} 
\providecommand{\DIFdelbegin}{} 
\providecommand{\DIFdelend}{} 
\providecommand{\DIFaddbeginFL}{} 
\providecommand{\DIFaddendFL}{} 
\providecommand{\DIFdelbeginFL}{} 
\providecommand{\DIFdelendFL}{} 
\newcommand{\DIFscaledelfig}{0.5}
\RequirePackage{settobox} 
\RequirePackage{letltxmacro} 
\newsavebox{\DIFdelgraphicsbox} 
\newlength{\DIFdelgraphicswidth} 
\newlength{\DIFdelgraphicsheight} 
\LetLtxMacro{\DIFOincludegraphics}{\includegraphics} 
\newcommand{\DIFaddincludegraphics}[2][]{{\color{blue}\fbox{\DIFOincludegraphics[#1]{#2}}}} 
\newcommand{\DIFdelincludegraphics}[2][]{
\sbox{\DIFdelgraphicsbox}{\DIFOincludegraphics[#1]{#2}}
\settoboxwidth{\DIFdelgraphicswidth}{\DIFdelgraphicsbox} 
\settoboxtotalheight{\DIFdelgraphicsheight}{\DIFdelgraphicsbox} 
\scalebox{\DIFscaledelfig}{
\parbox[b]{\DIFdelgraphicswidth}{\usebox{\DIFdelgraphicsbox}\\[-\baselineskip] \rule{\DIFdelgraphicswidth}{0em}}\llap{\resizebox{\DIFdelgraphicswidth}{\DIFdelgraphicsheight}{
\setlength{\unitlength}{\DIFdelgraphicswidth}
\begin{picture}(1,1)
\thicklines\linethickness{2pt} 
{\color[rgb]{1,0,0}\put(0,0){\framebox(1,1){}}}
{\color[rgb]{1,0,0}\put(0,0){\line( 1,1){1}}}
{\color[rgb]{1,0,0}\put(0,1){\line(1,-1){1}}}
\end{picture}
}\hspace*{3pt}}} 
} 
\LetLtxMacro{\DIFOaddbegin}{\DIFaddbegin} 
\LetLtxMacro{\DIFOaddend}{\DIFaddend} 
\LetLtxMacro{\DIFOdelbegin}{\DIFdelbegin} 
\LetLtxMacro{\DIFOdelend}{\DIFdelend} 
\DeclareRobustCommand{\DIFaddbegin}{\DIFOaddbegin \let\includegraphics\DIFaddincludegraphics} 
\DeclareRobustCommand{\DIFaddend}{\DIFOaddend \let\includegraphics\DIFOincludegraphics} 
\DeclareRobustCommand{\DIFdelbegin}{\DIFOdelbegin \let\includegraphics\DIFdelincludegraphics} 
\DeclareRobustCommand{\DIFdelend}{\DIFOaddend \let\includegraphics\DIFOincludegraphics} 
\LetLtxMacro{\DIFOaddbeginFL}{\DIFaddbeginFL} 
\LetLtxMacro{\DIFOaddendFL}{\DIFaddendFL} 
\LetLtxMacro{\DIFOdelbeginFL}{\DIFdelbeginFL} 
\LetLtxMacro{\DIFOdelendFL}{\DIFdelendFL} 
\DeclareRobustCommand{\DIFaddbeginFL}{\DIFOaddbeginFL \let\includegraphics\DIFaddincludegraphics} 
\DeclareRobustCommand{\DIFaddendFL}{\DIFOaddendFL \let\includegraphics\DIFOincludegraphics} 
\DeclareRobustCommand{\DIFdelbeginFL}{\DIFOdelbeginFL \let\includegraphics\DIFdelincludegraphics} 
\DeclareRobustCommand{\DIFdelendFL}{\DIFOaddendFL \let\includegraphics\DIFOincludegraphics} 

\newcommand{\Msun}{\mathrm{M}_\odot}

\newcommand{\epsplus}{\epsilon^{+}_{dis}}
\newcommand{\epsminus}{\epsilon^{-}_{dis}}

\usepackage{subfiles} 

\begin{document}

\title{\textit{NICER} neutron stars with dark energy and dark matter: effects on the inferred equation of state}

\author{Nathan Rutherford}
\email[Corresponding author: ]{nathan.rutherford@fqxi.org}
\affiliation{Department of Physics and Astronomy, University of New Hampshire, Durham, New Hampshire 03824, USA}
\affiliation{Foundational Questions Institute (FQxI), 235 Ponce de Leon Place, Suite M \#217 Decatur, GA 30030, USA}
\author{Chanda Prescod-Weinstein}
\email{chanda.prescod-weinstein@unh.edu}
\affiliation{Department of Physics and Astronomy, University of New Hampshire, Durham, New Hampshire 03824, USA}

\author{Anna~L.~Watts}
\email{A.L.Watts@uva.nl}
\affiliation{Anton Pannekoek Institute for Astronomy, University of Amsterdam, Science Park 904, 1098XH Amsterdam, the Netherlands}
\affiliation{Gravitation and Astroparticle Physics Amsterdam (GRAPPA), University of Amsterdam, 1098 XH Amsterdam, The Netherlands}

\bibliographystyle{unsrtnat}

\begin{abstract}
In this work, we compare neutron star configurations where dark energy or dark matter are present in the core, contrasting them against a baseline of pure baryonic stars with the same central energy density. From this comparison, we find that the dark energy models are more constrained by neutron star observations than the dark matter models, suggesting that neutron stars can place strong constraints on the dark energy parameter space. We consider four configurations: purely baryonic neutron stars, baryonic neutron stars admixed with bosonic or fermionic asymmetric dark matter (ADM) cores, and neutron stars with a dark energy core and baryonic shell, where the dark energy is described by the modified Chaplygin dark fluid (MCDF). The baryonic equation of state (EoS) is described by a piecewise polytropic (PP) EoS in the core, connected to results from chiral effective field theory up to 1.5 times nuclear saturation density. Using \textit{NICER} and radio pulsar timing measurements of the masses and radii of PSR~J0740$+$6620 and PSR~J0437$-$4715, as well as the \textit{NICER} mass-radius measurement of PSR~J0030$+$0451, we employ Bayesian inference to investigate how accounting for bosonic/fermionic ADM and MCDF cores affects the inferred EoS and neutron star properties. We find that MCDF cores, more than PP and ADM admixed models, substantially broaden neutron star mass-radius and pressure-energy density posterior distributions. Moreover, the MCDF EoS parameters can be tightly constrained by neutron star mass-radius observations. While bosonic and fermionic ADM cores decrease the inferred maximum mass, their posteriors strongly coincide with those of the PP model for ADM mass-fractions $F_\chi \leq 5\%$. These results demonstrate that current mass-radius observations cannot rule out MCDF cores, but they can constrain the MCDF parameter space to narrow regions. ADM admixed neutron stars, however, remain observationally identical to purely baryonic stars for all explored mass-fractions, even with improved mass-radius uncertainties. This suggests that mass-radius measurements alone are insufficient to detect ADM core inside neutron stars and that independent astrophysical or particle physics probes will be required.
\end{abstract}
\maketitle

\section{Introduction}
\label{sec:intro}
According to observations of anisotropies in the cosmic microwave background (CMB), the total mass-energy density of the Universe is dominated by two unknown substances, namely dark matter and dark energy. In particular, the Universe comprises about 68\% dark energy, 27\% dark matter, and only 5\% baryonic matter \cite{Planck2020}. Within the context of the standard model of cosmology, i.e., the $\Lambda$CDM model, the presence of cold dark matter can explain the rotation curves of galaxies \cite{Rubin1970,Rubin1978,Rubin1980}, the mass distribution of the Bullet cluster \cite{Clowe2006}, fluctuations in the CMB \cite{Planck2016,Planck2020}, and images of strongly gravitationally lensed systems \cite{Massey2010}. Moreover, the presence of dark energy, in the form a positive and non-vanishing cosmological constant $\Lambda$, provides excellent agreement with the observations of the accelerated expansion of the Universe \cite[see, e.g.,][]{Riess1998,SDSS:2005xqv,Bengaly2020}. However, despite its successes, the $\Lambda$CDM model faces a few challenges: we have yet to identify which dark matter candidate is the correct one; the Hubble tension is not resolved; and the cosmic coincidence problem and fine-tuning problems remain a challenge\cite{Zlatev:1998tr,Velten:2014nra,Padmanabhan:2002ji, Adams:2019kby}. While researchers explore a plethora of dark matter proposals, the Hubble tension, cosmic coincidence, and fining tuning all suggest that $\Lambda$ may be need to be replaced by a dynamical alternative. Such proposals include modified gravity theories, quintessence models, exotic dark fluids with positive energy density and negative pressures, extra dimensions, among others \cite[see][for a review]{Copeland:2006wr}. 

One dark energy model that has been proposed to address the issues with the cosmological constant $\Lambda$, is the Chaplygin dark fluid (CDF), which unifies dark matter and dark energy as a single fluid \cite{Kamenshchik:2001cp}. The equation of state (EoS) describing the CDF is given by $P = -A/\epsilon$, where $P$ is the pressure, $\epsilon$ is the energy density, and $A$ is a positive constant. The CDF is a well-motivated unified model of dark energy and dark matter because it simultaneously provides an explanation for the transition of dust-like matter in the early Universe to the current period of cosmic acceleration \cite{Xu2012}. However, the CDF was found to be in tension with multiple observations, such as gamma-ray bursts and Type Ia supernovae \cite[see][and references therein]{Zheng:2022vhj}. In order to address these issues, several variants of the CDF have been proposed in the literature, namely the generalized CDF ($P = -A/\epsilon^\gamma$, where $\gamma \in (0,1]$) \cite{Bento:2002ps}, the modified CDF ($P = A\epsilon - B/\epsilon^\gamma$, where $A$ and $B$ are positive constants) \cite{Debnath:2004cd}, and the so-called new generalized CDF ($P = -A(a)/\epsilon^\gamma$, where $a$ is the cosmological scale factor) \cite{Zhang:2004gc}. Using high-redshift quasar data, \cite{Zheng:2022vhj} showed that the modified CDF (MCDF) strongly agrees with the observations, while the generalized CDF and the new generalized CDF cannot be distinguished from the $\Lambda$CDM.

If dark energy is evenly distributed throughout the Universe as observations suggest, then it could be inside the cores of neutron stars and affect their measurable properties \cite{Rahaman:2010mr,Bhar:2016nlq,Bhar:2021qti,Panotopoulos:2021dtu, Tello-Ortiz:2020svg,Estevez-Delgado2021,Pretel:2023nhf,Araujo:2024txe,Pretel:2024tjw,Pretel:2024vvt,Pretel:2024a, Araujo:2025tlv}. Dark energy can generate two types of stars: neutron star-like objects in which the star is described as the EoS of the assumed dark energy model \cite{Rahaman:2010mr, Bhar:2016nlq,Tello-Ortiz:2020svg,Bhar:2021qti, Panotopoulos:2021dtu,Pretel:2023nhf} and a hybrid star with a dark energy core and a baryonic matter shell \cite{Araujo:2024txe,Pretel:2024a,Pretel:2024vvt,Pretel:2024tjw}. Both types of stars have been shown to produce maximum masses above 2$\Msun$ with radii between 10-14 km, thus satisfying the current neutron star observational data from radio pulsar measurements of \cite{Demorest:2010bx,Antoniadis13,Fonseca16,Arzoumanian18, Cromartie20, Fonseca21,Shamohammadi23} and NASA's Neutron Star Interior Composition Explorer (\textit{NICER}; \cite{Gendreau16}) \cite{Pretel:2024vvt,Pretel:2024a,Araujo:2024txe}. Although purely dark energy stars have been shown to satisfy the \textit{NICER} data, such stars are not directly observable by \textit{NICER}. This leaves open the question of how observations by \textit{NICER} and similar instruments may provide insights into the hybrid scenario, where there is both a dark energy core and also a baryonic matter shell.\footnote{Recently, \citet{Araujo:2025tlv} showed that neutron stars can be modeled as three component stars of baryonic matter, dark energy, and dark matter, and simultaneously match the data provided by \textit{NICER} and LIGO/Virgo/KAGRA. Although the configurations discussed in \citet{Araujo:2025tlv} are physically interesting, we will leave analyzing these systems for future work.}

In contrast to the recent literature on the affects of MCDF cores on neutron star properties, there is a substantial body of literature suggesting that dark matter, too, can accumulate inside neutron stars and affect their masses, radii, and tidal deformabilities \cite[see, e.g.,][]{Bertone:2007ae,Kouvaris:2010jy,Kouvaris:2011fi,Bramante:2013hn,Ellis2018,Nelson2019,Das2019,Das2020,Ivanytskyi2020,Sen2021,Collier2022,Miao_2022,Karkevandi2022,Routaray2023,Rutherford23,Sagun2023,Bramante:2023djs,Giangrandi2024,Rutherford:2024bli, Guha2024,Hajkarim:2024ecp, Mariani2024, Shawqi:2024jmk,Shawqi:2025cca,Thakur2024a, Thakur2024b, Karkevandi2024}. There are two spatial regimes in which dark matter can accumulate: inside the core of the neutron star interior and a dark matter halo that extends through and beyond the baryonic surface. In a dark matter core configuration, the neutron star mass, radius, and tidal deformabilities are reduced compared to a star with an identical baryonic central energy density \cite{Ellis2018,Das2021,Karkevandi2022,Collier2022,Bastero-Gil:2024kjo,Giangrandi2022, Konstantinou2024}. Generally, if dark matter forms a halo, neutron star masses and tidal deformabilities increase \cite{Nelson2019,Ivanytskyi2020, Sagun2022,Miao_2022, Guha2024}. If the dark matter halo is sufficiently compact, it can mimic a dark matter core and reduce the masses and radii of these stars \cite{Shawqi:2024jmk}. More recently, \citet{Shawqi:2025cca} investigated the effects of rotation on dark matter halo distributions and found that rapid rotation of the baryonic portion of the neutron star reduces the dark matter halo radius compared to the same non-rotating neutron star. Beyond these structural effects, the presence of a dark matter halo also modifies the exterior spacetime and alters the trajectories of photons emitted from the neutron star surface, thereby changing the interpretation of \textit{NICER} measurements \cite{Miao_2022,Shakeri2024, Shawqi:2024jmk, Shawqi:2025cca}. Since dark matter can affect the measurable properties of neutron stars, the possible presence of dark matter, along with the presence of dark energy cores, should be accounted for in analyses of neutron stars.

To investigate the structural impact of dark matter and dark energy cores within neutron star interiors, we need mass and radius measurements of neutron stars. This is because different interior compositions, whether purely baryonic, admixed with dark matter, or containing a dark energy core, predict distinct mass-radius relations, making these measurements a strong probe of the neutron star EoS. Mass and radius measurements can therefore constrain which EoS models are consistent with the observed neutron stars, and in turn place bounds on the properties of a possible dark matter or dark energy component present in the interior.

In practice, the \textit{NICER} and radio timing measurements are used to constrain the neutron star mass-radius relation, which is the relation of all possible, stable neutron star masses and radii corresponding to an EoS \cite{Lindblom1992}. The dense matter EoS describes the relationship of pressure as a function of energy density throughout the star and theoretically encodes the microphysical behavior of the matter inside the stellar interior. By computing the mass-radius relation for a specific EoS and comparing it to the \textit{NICER} measurements, it is possible to constrain the dense matter EoS. In order to derive the masses and radii of neutron stars, \textit{NICER} employs Pulse Profile Modeling \cite[PPM, see][]{Watts19a,Bogdanov19a,Bogdanov19b,Bogdanov21}, the relativistic ray-tracing Bayesian inference technique that exploits the thermal emission from hot spots on the surface of millisecond pulsars. The PPM analysis delivers the posterior probability distributions of not only the masses and radii of neutron stars, but also other important parameters, such as hot spot geometry, temperature, and inclination. To date, \textit{NICER} has delivered PPM derived mass-radius posteriors for five millisecond pulsars: PSR J0740$+$6620 \cite[PSR~J0740 from hereon;][]{Wolff21,Fonseca21,Riley21,Miller21,Salmi22,Salmi23,Salmi24,Dittmann24,Hoogkamer:2025ype}, PSR J0030$+$0451 \cite[PSR~J0030 from hereon;][]{Riley19,Miller19,Salmi23,Vinciguerra23,Vinciguerra24,Kini26}, PSR J0437$-$4715 \cite[PSR~J0437 from hereon;][]{Choudhury24,Reardon24, Miller26}, PSR~J1231$-$1411 \cite[PSR~J1231 from hereon;][]{Salmi:2024bss, Qi:2025mpn}, PSR~J0614$-$3329 \cite[PSR~J0614 from hereon;][]{Reardon:2023zen, Miles:2024rjc, Mauviard:2025dmd, Miller:2026vpr}, PSR~J1614$-$2230 \cite[PSR~J1614 from hereon;][]{NANOGrav:2023hde,Mauviard:2026gzc}, and PSR~J2124$-$3358 \citep[J2124 from hereon;][]{Gonzalez-Caniulef:2026wcm}\footnote{Since the constraints for PSR~J1231 and PSR~J2124 are relatively weak compared to those of the other sources, it will be excluded from the remainder of this analysis. Furthermore, since \citet{Mauviard:2025dmd} showed that the inferred neutron star mass-radius relation and the resulting EoS constraints remains consistent with prior analyses to within one standard deviation regardless of the inclusion of PSR~J0614. Consequently, we exclude this source from this analysis. Finally, during the writing of this manuscript, the PPM-derived posteriors \cite{Mauviard:2026gzc} and subsequent EOS analysis \cite{Mendes:2026svw} of PSR~J1614 were published. Although PSR~J1614's high gravitational mass would be interesting to include in this analysis, we leave including this source to future work due to the high computational cost of re-performing our Bayesian inferences.} These mass-radius measurements have been extensively investigated in EoS studies concerning only baryonic matter \cite[see, e.g.,][]{Raaijmakers19,Miller19, Raaijmakers20,Miller21,Raaijmakers21,Legred21,Biswas22,Huth22,Annala23,Huang:2023grj,Pang24, Rutherford:2024srk,Li:2024sft,Huang:2024rvj, Kurkela24, Mauviard:2025dmd, Mendes:2026mgc,Mendes:2026svw} and also dark matter admixed neutron stars \cite[see, e.g.,][]{Das2022a,Miao_2022,Rutherford23,Rutherford:2024bli, Wei:2026gcc, Hu:2025eyt, Shahrbaf:2025hsw}.

Two works, \citet{Pretel:2024vvt} and \citet{Araujo:2024txe}, have used the derived \textit{NICER} PPM mass-radius measurements and gravitational wave measurements from LIGO/VIRGO in the context of dark energy cores inside neutron star interiors. \citet{Araujo:2024txe} modeled neutron stars with dark energy cores as an admixed system, where the vacuum energy EoS ($p = - \epsilon$) is used for the dark energy component and several tabulated nuclear EoS models for the baryonic component. To characterize the impact of the vacuum energy EoS, the authors defined the energy density fraction $y$, which described the ratio of the baryonic energy density to summed energy densities of dark energy and baryonic matter. Using these EoS models in an admixed system and the energy density fraction, the authors generated neutron star mass-radius and mass-tidal deformabilities, and compared them to the measurements of PSR~J0740, PSR~J0030, GW170817 \cite{gw170817}, GW190425 \cite{gw190425}, PSR~J0348$+$0432 \cite{Antoniadis13}, and PSR~J1614$+$2230 \cite{Demorest:2010bx}. The results of \citet{Araujo:2024txe} showed increasing $y$ reduces neutron star masses, radii, and tidal deformabilities. In addition, the authors placed limits on the maximum value of $y$ for several nuclear EoS models. In \citet{Pretel:2024vvt}, the authors consider neutron stars with a dark energy core surrounded by a baryonic matter shell, where the transition from the core to shell was described using a rapid first-order phase transition. Here, they model the core using the MCDF EoS with $\gamma = 1$ and two physics motivated nuclear EoS models. The authors then compared the effects of the MCDF EoS parameters, namely the barotropic pressure constant $A$ and phase transition jump parameter $\alpha$, to the measurements of PSR~J0740, PSR~J0030, and GW170817. \citet{Pretel:2024vvt} found that variations of $\alpha$ and $A$ can significantly impact the properties of neutron stars and remain consistent with observations. Here, both studies investigated the impact of dark energy cores on neutron star properties and determined that they can remain in agreement with observations.

In this paper we investigate how to use Bayesian inference to constrain MCDF cores and compare those inferences to ones for neutron stars made of purely baryonic matter and inferences for dark matter admixed cores. We assume the dark energy core is modeled as the MCDF EoS considered in \citet{Pretel:2024vvt}. In order to construct the dark matter cores, we assume the dark matter could be described by either the \textit{bosonic} or \textit{fermionic} asymmetric (ADM) models developed in \citet{Nelson2019} and investigated in \citet{Rutherford23,Rutherford:2024bli}. Since the baryonic physics inside neutron star interiors is not fully understood at the nuclear level, we adopt the \citet{Hebeler2013} piecewise polytropic model and the \citet{Keller2023} chiral effective field theory calculations to characterize baryonic matter and account for the uncertainties associated with the baryonic EoS. We consider the derived \textit{NICER} mass-radius measurements of PSR~J0740, PSR~J0030, PSR~J0437 reported in \citet{Salmi24}, \citet{Vinciguerra24}, and \citet{Choudhury24}, respectively. One of the main objectives of this work is to explore the effects of dark energy cores on the inferred neutron star properties and compare them to inferences of neutron stars comprised of baryonic matter and admixed with ADM, respectively. 

In addition to the main objective on the effects of MCDF cores on inferred neutron star properties, this work has two other objectives: the first is to determine the promise of constraining the MCDF EoS parameters using \textit{NICER} neutron star measurements and the other is to investigate if ADM admixed neutron stars remain indistinguishable from purely baryonic neutron stars for high ADM mass-fractions, thus extending the results of our previous work, \citet{Rutherford:2024bli}.

The work presented here shows that neutron stars with MCDF cores substantially impact the dense matter EoS \textit{a priori} and \textit{a posteriori} when compared to the PP model and ADM admixed models. In particular, both the prior and posterior inferences reveal that neutron stars with MCDF cores broaden the dense matter EoS pressures at intermediate densities and prefer substantially higher maximum masses compared to that of the PP model and ADM admixed models. Moreover, this work shows \textit{NICER} mass-radius measurements can place stringent constraints on the MCDF EoS parameters, namely the barotropic pressure constant and the phase transition density. Finally, in \citet{Rutherford:2024bli}, we performed a Bayesian analysis on neutron stars admixed with fermionic ADM cores, and found that the posteriors accounting for fermionic ADM were indistinguishable from those that neglected fermionic ADM for mass-fraction below 1.7\% and mass-radius observational uncertainties at or above the 2\% level. Expanding on this from our prior work in \citet{Rutherford:2024bli}, we find that ADM admixed neutron stars, regardless of whether they are bosonic or fermionic in nature, are as equally consistent with the \textit{NICER} data as purely baryonic stars for ADM mass-fractions below 5\%.

The remainder of this work is organized as follows. In Sec.~\ref{sec:eos models}, we define the MCDF, bosonic and fermionic ADM, and baryonic EoS models. In Sec.~\ref{sec: modeling}, we describe the numerical approach to computing neutron star properties with ADM and MCDF cores using the two-fluid TOV and single-fluid TOV equations, respectively. Sec.~\ref{sec:inference framework} discusses our Bayesian parameter estimation framework for computing the inferred dense matter EoS. In Sec.~\ref{sec: priors and source selection}, we define the PP EoS priors, the bosonic and fermionic ADM EoS priors, MCDF EoS priors, and the considered \textit{NICER} mass-radius measurements. In Sec~\ref{sec:results}, we 
 first discuss the impact of fermionic ADM, bosonic ADM, and MCDF cores on the EoS priors relative to the PP model priors, and then turn to their respective effects on the inferred dense EoS and neutron star properties. Sec.~\ref{sec:results} also addresses the constraints on the MCDF EoS parameters. Lastly, we summarize our results and conclude in Sec.~\ref{sec:summary and conclusions}.

\section{EoS models}\label{sec:eos models}
In this work, we examine four distinct configurations of neutron stars: ones made entirely of baryonic matter, ones made of an admixture of both ADM and baryonic matter, and ones featuring a dark energy core surrounded by a shell of baryonic matter. In this section, we describe the EoSs of baryonic matter, the \citet{Nelson2019} ADM model, and the MCDF model employed to generate these neutron star models.

\subsection{Baryonic EoS}\label{subsec:baryonic eos}
In order to capture the uncertainties associated with the baryonic matter EoS, we need to make prior assumptions over all plausible densities. Therefore, we use the piecewise polytropic (PP) high-density EoS extension of \citet{Hebeler2013}, which describes three polytropes divided by two varying transition densities, to span the range of possible EoSs beyond $1.5 n_0$, where $n_0 = 0.16 \, \text{fm}^{-3}$ is the nuclear saturation density. The value of $1.5 n_0$ marks the approximate upper boundary of validity for chiral effective field theory ($\chi$EFT) calculations, beyond which the theoretical uncertainties grow rapidly and are too large reliably constrain the dense matter EoS, thus necessitating a more parametric high-density extension \cite{Hebeler2013,Tews:2018kmu,Drischler2021ARNPS}. We employ the PP model to parametrize baryonic microphysics because its parameterized nature allows us to explore a broad range of the physically permissible mass-radius plane while remaining versatile enough to fit a wide variety of tabulated, physics-motivated EoS models \cite{Read2009,Hebeler2013,Greif19}, meaning that when adequately sampled it is capable of faithfully capturing the full uncertainties of the baryonic EoS. This flexibility, combined with the integration of $\chi$EFT results at lower densities to ensure physical accuracy in that regime, makes the PP model a well-motivated and reliable framework for parameterizing baryonic microphysics across all relevant densities. 

At densities below $\approx 0.5 n_0$, we consider the Baym-Pethick-Sutherland (BPS) crust EoS \cite{Baym71}. The BPS model is considered here to describe the outer and inner crust, providing a consistent thermodynamic baseline before reaching the density regime where $\chi$EFt becomes the more appropriate framework describing the bulk nucleon interaction. We then connect to the $\chi$EFT calculations of \citet{Keller2023} to capture the uncertainties of the baryonic EoS up to $1.5 n_0$. We choose the \citet{Keller2023} calculations specifically because they represent one of the most up-to-date next-to-next-to-next leading order (N$^3$LO) $\chi$EFT frameworks and provide a well quantified uncertainty band over the density range in which $\chi$EFT calculations are relevant. To model the \citet{Keller2023} $\chi$EFT framework, we follow our prior work \citet{Rutherford:2024srk}, and fit a single polytrope to the \citet{Keller2023} N$^3$LO $\chi$EFT calculations for densities ranging from $0.5n_0$ to $1.5n_0$. Within the density range of $[0.5n_0, 1.5n_0]$, we find that the \citet{Keller2023} $\chi$EFT band is well fitted by $K_{\chi \rm{EFT}} \in [2.207, 3.056] \, \mathrm{MeV \, fm^{-3}}$, $\Gamma_{\chi \rm{EFT}} \in [2.361, 2.814]$, where $K_{\chi \rm{EFT}}$ and $\Gamma_{\chi \rm{EFT}}$ are the matching constant and adiabatic index, respectively. 

\subsection{Bosonic \& fermionic ADM model}\label{subsec:ADM eos}
The ADM model is motivated by the observation that the mass density of dark matter in the Universe is about five times that of baryonic matter, suggesting deep ties in the cosmic history of dark matter and baryons \cite{Zurek2013,Petraki2013}. In the ADM paradigm, the connection between the baryon and dark matter mass densities arises naturally when an asymmetry between the dark matter and anti-dark matter number densities is connected to the baryon asymmetry in the early Universe. Therefore, the dark matter density is determined by its own asymmetry as opposed to being set by its annihilation cross-section. A ``dark asymmetry" like this would allow for significant repulsive self-interactions and small attractive interactions with baryons. In addition, significant ADM self-interactions could address some of the tensions between the predictions of collisionless cold dark matter and observations on galactic and sub-galactic scales \cite[see][and references therein]{Petraki2013}.

The \citet{Nelson2019} ADM model describes a MeV-GeV mass-scale bosonic or fermionic dark matter particle with repulsive self-interactions mediated by a eV-MeV mass-scale vector gauge boson, $\phi_\mu$. Repulsive interactions are necessary to stabilize the bosonic dark matter structure against gravitational collapse, whereas in the fermionic case the Fermi degeneracy pressure provides the necessary stabilization against gravity. These differing constructions motivate a separate treatment between the bosonic and fermionic cases, rather than considering one in this analysis. This is because these two stabilization mechanisms are fundamentally different, the bosonic and fermionic ADM cases have differing allowed ADM parameter spaces because the self-repulsion is allowed to be zero in the fermionic case, which in-turn also expands the allowed fermionic ADM particle mass compared to that of the bosonic case \cite[see, Sec.~\ref{sec: priors and source selection} and][]{Rutherford23,Rutherford:2024bli}. In addition, the vector gauge boson also carries the Standard Model baryon number, which is needed to create a ``dark asymmetry'' in the early Universe.

By assuming that all of the ADM particles have thermally equilibrated with the baryons inside the neutron star, the ADM EoS can be computed at zero temperature. Furthermore, given that the ADM particles are in the dense baryonic background of the neutron star, the mean-field approximation can be employed. Within the mean-field approximation framework,s given by  the \citet{Nelson2019} bosonic ADM EoS, in terms of $\hbar$ and $c$, is given by
\begin{align}
    \epsilon_\chi &= m_\chi c^2 n_\chi + \frac{g_\chi^2}{2 m_\phi^2} \frac{\hbar^3}{c} n_\chi^2 \label{bosonicADM_eps}\\
    \nonumber \\
    P_\chi &= \frac{g_\chi^2}{2 m_\phi^2} \frac{\hbar^3}{c} n_\chi^2 \label{bosonicADM_pres},
\end{align}
where $\epsilon_\chi$ is the ADM energy density, $P_\chi$ is the ADM pressure, and $n_\chi$ is the ADM number density. While the \citet{Nelson2019} fermionic ADM EoS, in units of $\hbar$ and $c$ restored, is 
\begin{multline}\label{fermionicADM_eps}
     \epsilon_\chi = \frac{c^5 m_\chi^4}{8 \pi^2 \hbar^3} \Big[\sqrt{1+z^2} (2z^3 +z) \\- ln(z + \sqrt{1+z^2})   \Big]  + \frac{g_\chi^2}{2 m_\phi^2} \frac{c^5 (m_\chi z)^6}{\hbar^3 (3\pi^2)^2}
\end{multline}
\begin{multline}\label{fermionicADM_pres}
     P_\chi = \frac{c^5 m_\chi^4}{8 \pi^2 \hbar^3} \Big[\sqrt{1+z^2} \Big(\frac{2}{3}z^3 -z \Big) \\+ ln(z + \sqrt{1+z^2})   \Big] + \frac{g_\chi^2}{2 m_\phi^2} \frac{c^5 (m_\chi z)^6}{\hbar^3 (3\pi^2)^2},
\end{multline}
where $z = \hbar k_\chi/m_\chi c$ is the relativity parameter defined in terms of the ADM Fermi momentum $k_\chi = \left(3\pi^2 n_\chi \right)^{1/3}$. 

\subsection{Modified Chaplygin dark fluid EoS}\label{subsec:cdf eos}
If the Universe is uniformly distributed with a dark-energy fluid (e.g., a Chaplygin-like gas), neutron stars could contain a dark-energy component within their cores. Because their extreme internal conditions can theoretically support dark energy fluid cores, neutron stars serve as ideal laboratories for probing dark energy. The MCDF EoS is a compelling candidate for this component, as it possesses a negative pressure term characteristic of dark energy alongside a barotropic correction that ensures hydrostatic stability. Recent work by \citet{Pretel:2024tjw} confirmed that neutron stars with MCDF cores are stable under radial perturbations and are consistent with NICER and GW170817 constraints, justifying further investigation into these configurations using Bayesian inference. To model this dark energy component, we employ the MCDF EoS, defined as

\begin{equation}\label{CDF_eos}
P_{DE}(\epsilon_{DE}) = A \epsilon_{DE} - \frac{B}{\epsilon_{DE}},
\end{equation}
where $P_{DE}$ is the MCDF pressure, $\epsilon_{DE}$ is the MCDF energy density, $A$ is a positive dimensionless constant, $B$ is positive constant with cgs units $[\mathrm{g^2 cm^{-2} s^{-4}}]$. The first term describes a barotropic fluid that ensures the dark energy core can sustain positive pressures, while the second represents the original Chaplygin gas EoS and produces the negative pressure responsible for cosmic acceleration.

Modeling a neutron star with a dark energy core and a baryonic shell requires specifying how these two distinct, non-mixing fluids meet at their boundary\footnote{See \cite{Pretel:2024vvt,Pretel:2024a,Pretel:2024tjw, Araujo:2024txe,Araujo:2025tlv} for similar treatments of dark energy fluids inside neutron star cores.}. To define an EoS with two distinct fluids, which do not mix nor occupy the same spatial region, we construct the stellar interior using a sharp interface (i.e., the Maxwell construction) between the two fluids, enforcing pressure continuity while also permitting a discontinuous energy density jump. This is the physically appropriate approximation for the non-interacting, disjoint two-fluid configuration considered here because the other construction typically used for first-order phase transitions, namely the Gibbs construction, requires that the two fluids share conserved charges and can exchange particles. These two conditions are not obviously satisfied when considering a dark energy fluid with no well-defined particle interpretation\footnote{See \cite{Essick:2023fso, Han:2019bub} for further details on both the Maxwell and Gibbs constructions.}. Thus, under the Maxwell construction, the total neutron star EoS is given by

\begin{equation}\label{DE_stellar_fluid}
P(\epsilon) = 
\Bigg\{
    \begin{array}{lr}
        A \epsilon_{DE} - \frac{B}{\epsilon_{DE}}, & 0 \leq r \leq R_{dis}\\
        P_B(\epsilon_B), & R_{dis} \leq r \leq R,
    \end{array}
\end{equation}
where $r$ is the radius coordinate, $R_{dis}$ is the radius in which the transition from dark energy to baryonic matter occurs. Since the pressures on either side of $R_{dis}$ must be the same to ensure pressure continuity, Eq.~\ref{DE_stellar_fluid} implies that the $B$ parameter can be written as

\begin{equation}\label{B_param}
    B = A (\epsilon^{+}_{dis})^2 - \epsilon^{+}_{dis} P_{B}(\epsilon^{-}_{dis}),
\end{equation}
where $\epsplus$ is the energy density defining the boundary of the MCDF core, $\epsminus$ is the energy density where the baryonic shell begins, $P_B(\epsminus)$ is the baryonic pressure evaluated on the lower side of the transition. Rather than sampling $\epsplus$ and $\epsminus$ independently (and enforcing the constraint $\epsminus \leq \epsplus$), we define the energy density jump parameter $\alpha = \epsminus/\epsplus \leq 1$ \cite{Pretel:2024tjw}. This $\alpha$ parameter reduces the transition into a single dimensionless quantity that directly characterizes the strength of the transition from the dark energy core to the baryonic shell. For a fixed value of $\epsplus$, $\alpha = 1$ and corresponds to a continuous transition with no density jump, while $\alpha$ close to zero would correspond to a larger energy density jump. This parameterization is particularly convenient for Bayesian inference because $\alpha \in (0,1]$ is bounded and physically interpretable. By combining Eq.~\ref{CDF_eos}, Eq.~\ref{B_param}, and the definition of the energy density jump variable $\alpha$, the free parameters of the MCDF model are $A$, $\epsplus$, and $\alpha$.  

\section{Modeling neutron star properties: ADM vs. dark energy}\label{sec: modeling}
In order to compute the masses and radii of neutron stars admixed with ADM, we employ the two-fluid formalism, which assumes the dominant interaction between the Standard Model and ADM is gravitational \cite{Sandin2009}. However, the global properties of neutron stars with a dark energy core can be computed by modeling the transition from the core to the shell using the Maxwell Construction and solving the traditional Tolman-Oppenheimer-Volkoff (TOV) equations \cite{Pretel:2024vvt,Pretel:2024a}. In this section, we outline how to compute the mass-radius relation of neutron stars admixed with ADM as well as stars with cores comprised of dark energy.

\subsection{ADM admixed neutron stars and the two-fluid TOV equations}\label{subsec: two-fluid}
The mass-radius relation of ADM admixed neutron stars can be computed by solving the two-fluid TOV equations, which are a direct consequence of the two-fluid formalism\footnote{Although the choice of the two-fluid formalism is not fully consistent with the ADM paradigm since it neglects any non-gravitational interaction between baryonic matter and ADM, the coupling between the Standard Model and ADM is expected to be small compared to the coupling of ADM to itself (see Sec.~\ref{sec:eos models} and \citet{Rutherford:2024bli} for further details). Additionally, since this work uses the PP EoS of \citet{Hebeler2013} to model the baryonic EoS, the conserved current $J_\mu^B$ cannot be computed because the \citet{Hebeler2013} model is parameterized, thus there is no corresponding Lagrangian. Therefore, since this considers the \citet{Hebeler2013} PP model to capture the uncertainties of the baryonic EoS and we assume $g_B \ll g_\chi$, the two-fluid formalism is an appropriate framework to compute the properties of ADM admixed neutron stars. However, this can be improved by using the interacting two-fluid model of \citet{Hajkarim:2024ecp}, which we leave for future work.}, and simultaneously varying the central densities of ADM and baryonic matter given an EoS for both components \cite[see][and references therein]{Ellis2018, Nelson2019,Das2020,Das2021, Karkevandi2022,Giangrandi2023,Konstantinou2024}. The two-fluid TOV equations can derived from the standard single-fluid TOV equations using the two-fluid formalism. By assuming the dominant fluid interaction is gravitational, the two-fluid formalism implies that ADM and baryonic matter each satisfy their own conservation of energy-momentum equation (i.e,, $\nabla_\mu T^{\mu \nu} = 0$) since. Conservation of energy-momentum for both ADM and baryonic matter is equivalent to
\begin{align}
    P &= P_B + P_{\chi}\label{pB_p_ADM}\\
    \epsilon &= \epsilon_B + \epsilon_\chi \label{epsB_epsADM},
\end{align}
where $P$($\epsilon$) is the total pressure (energy density). By substituting $P$ and $\epsilon$ into the TOV equations, we have
\begin{align}
      &\frac{dP_B}{dr} = - \left( \epsilon_B +P_B \right) \frac{Gc^2M + 4\pi r^3 G P}{c^2 r \left[rc^2-2GM  \right] } \label{dpbdr}\\
       &\frac{dP_\chi}{dr} = - \left( \epsilon_\chi +P_\chi \right) \frac{Gc^2M + 4\pi r^3 G P}{c^2 r \left[rc^2-2GM  \right] } \label{dpchidr}\\
        &\frac{dM_B}{dr} = 4\pi r^2 \frac{\epsilon_B}{c^2} \\ 
         &\frac{dM_\chi}{dr} = 4\pi r^2 \frac{\epsilon_\chi}{c^2},
\end{align}
where $M_\chi$ is the gravitational mass of ADM, $M_B$ is the gravitational mass of baryonic matter, and $M = M_B+M_\chi$. Given the ADM EoS and baryonic EoS, the two-fluid TOV equations are numerically integrated radially outward for a range of pairs of baryonic and ADM central energy densities until one of the fluid pressures reaches zero, marking either a neutron star with an ADM core if $P_\chi = 0, \, P_B \neq 0$ or the baryonic surface of a star with an ADM halo if $P_\chi \neq 0, \, P_B = 0$. Furthermore, the radial values at which $P_\chi = 0$ and $P_B = 0$ define the ADM radius ($R_\chi$) and Baryonic radius ($R_B$), respectively. Thus, the percent ratio of the total ADM gravitational mass to the total gravitational mass of the neutron star, i.e. the ADM mass-fraction, can then be defined as
\begin{equation}
    F_\chi = \frac{M_\chi(R_\chi}{M_\chi(R_\chi) + M_B(R_B)}*100.
\end{equation}
Defining $F_\chi$ is important to the analysis of ADM admixed neutron stars because most ADM accumulation mechanisms, such neutron bremsstrahlung \cite{Nelson2019}, accretion \cite{Ivanytskyi2020}, and neutron decay to ADM \cite{Ellis2018}, compute the total accumulated ADM mass in terms of a fraction of $M_{NS}$, the total neutron star mass. This makes the ADS mass fraction a physically interesting parameter to consider. Lastly, in order to compute a desired $F_\chi$ for a given ADM and baryonic matter EoS using the two-fluid TOV equations, we follow the approach outlined in \citet{Rutherford23}. The \citet{Rutherford23} numerical algorithm is implemented in the \texttt{NEoST v2.2.0} neutron star EoS inference code \cite{Raaijmakers:2025hbz}\footnote{\url{https://github.com/xpsi-group/neost}}.

\subsection{Neutron Stars with a dark energy core}\label{subsec: single-fluid TOV}
To investigate the impact of an MCDF dark energy core on neutron star observables, and by extension the mass-radius relation, we model the compact object as a hybrid star. A hybrid star is a neutron star comprised of two distinct regions, which are spatially separated by a sharp boundary where the constituent fluids do not mix. Typically, these are neutron star modeled with a quark core surrounded by a shell of nuclear matter \cite[see, e.g., ][]{Zhao:2020dvu,Han:2019bub, Chatziioannou:2019yko,Annala:2019puf}. However, this work considers hybrid stars with a dark energy core described by the MCDF EoS, surrounded by a baryonic matter shell defined by the high-density PP EoS model. 

This configuration fundamentally differs from the ADM scenario detailed in Sec.~\ref{subsec: two-fluid}. In ADM admixed neutron stars, the ADM and baryonic matter co-occupy the same spatial volume and interact primarily through gravity, which necessitates using the two-fluid TOV equations to simultaneously solve for their overlapping structure. On the other hand, since the MCDF core and baryonic shell in our hybrid star model do not spatially overlap, and are governed by a sharp transition from one fluid to the next, the macroscopic neutron star structure can be determined by integrating a single pressure profile. This is most clearly shown in Eq.~\ref{DE_stellar_fluid}. Therefore, the hybrid neutron star masses and radii can be computed using one central energy density and the standard single-fluid TOV equations \cite{Tolman1939,Oppenheimer1939}. 

For a given EoS, the single-fluid TOV equations are solved by integrating radially outward until the fluid pressure vanishes. For a hybrid star defined by the Maxwell construction, such as that of Eq.~\ref{DE_stellar_fluid}, the integration of the single-fluid TOV equations begins with using the first fluid's EoS until it is halted at the transition energy density $\epsplus$ corresponding to $R_{\rm dis}$. Then, depending on the width of the density jump (which is determined by the parameter $\alpha$) from the first fluid in the core to the surrounding shell of the second fluid, the integration is resumed using the EoS of the second fluid at the pressure in which the integration was halted, but at the lower energy density $\epsminus$. Integrating across a range of central energy densities in this way produces the full mass-radius relation for these hybrid stellar configurations.

One consequence of computing a piecewise transition produced by dark energy cores, but also the formation of quark matter cores, in the TOV equations is the generation of two separate stable mass-radius relations (i.e., ``branches''): a primary branch describing hybrid stars with an MCDF core and baryonic shell, and a secondary branch at higher radii consisting of purely baryonic stars. Mass-radius curves with these two stable branches can produce neutron stars with the same masses, but different radii, which are known as twin stars \cite[see, e.g.,][]{Blaschke:2020qqj,Christian:2021uhd,Essick:2023fso,Jimenez:2024hib, Li:2024sft}. Although such configurations are physically possible, the secondary branch introduces significant computational computational challenges during the Bayesian inference process. Specifically, because a single neutron star mass can correspond to multiple radii in the twin star case, transforming the posteriors across a multi-valued mass-radius relation is computationally difficult. Furthermore, the inference becomes confounded since the second branch, which in this context contains purely baryonic matter, remains mathematically defined by a combination of both the MCDF and PP EoS parameters. As a result, given these computational challenges, and because this work focuses specifically on neutron stars containing an MCDF core, we discard the second branch and leave a more complete Bayesian analysis of twin stars for future work.

\section{Inference framework}\label{sec:inference framework}
In this work, we follow the analysis framework used in \citet{Rutherford23,Rutherford:2024bli}, which is accounts for the possible presence of an ADM component inside the neutron star interior, and is implemented in the open-source EoS inference code \texttt{NEoST v2.2.0} \cite{Raaijmakers:2025hbz}. To account for the MCDF EoS model, modifications to the \texttt{NEoST} code were required, which can be found in \texttt{NEoST v3.0.0}\footnote{\texttt{NEoST v3.0.0} is currently an unofficially released version of NEoST, thus please see the \texttt{Dark\_Energy\_Adaptation} GitHub branch at \url{https://github.com/xpsi-group/neost/tree/Dark_Energy_Adaptation}.}. Furthermore, a full reproduction package, including the posterior samples and scripts to generate the data and plots in this work, will be available
in a Zenodo repository at \cite{Rutherford2026_dataset}. Here, we briefly summarize the \citet{Rutherford23} method, outline the modifications required to account for the MCDF model, and state all imposed constraints.

To begin, we apply Bayes' theorem to establish a posterior distribution on all EoS parameters with the following form:

\begin{align}
    p(\boldsymbol{\theta}, \boldsymbol{\epsilon_c} |\mathbf{d})\nonumber &\propto p(\boldsymbol{\theta}) p(\boldsymbol{\epsilon_c}|\boldsymbol{\theta}) ~p(\mathbf{d} |\boldsymbol{\theta}, \boldsymbol{\epsilon_c}) \\
    & \propto p(\boldsymbol{\theta})  p(\boldsymbol{\epsilon_c}|\boldsymbol{\theta}) p\boldsymbol{(}\mathbf{d}|\mathbf{M}(\boldsymbol{\theta},\boldsymbol{\epsilon_c}), \mathbf{R}(\boldsymbol{\theta},\boldsymbol{\epsilon_c})\boldsymbol{)} \label{eq:Bayes step 1}, 
\end{align}
where $\boldsymbol{\theta}$ is the vector containing all EoS parameters, $\boldsymbol{\epsilon_c}$ is the vector containing the central energy densities, $\mathbf{d}$ is the vector containing the masses and radii of the sources from each scenario, $\mathbf{M}(\boldsymbol{\theta},\boldsymbol{\epsilon_c})$ is the mass of a produced neutron star, and $\mathbf{R}(\boldsymbol{\theta},\boldsymbol{\epsilon_c})$ is the radius. Here, we sample over $F_\chi$ for ADM admixed neutron stars because the ADM mass-fraction is a function of central energy densities of both baryonic matter and ADM. Thus, we define $F_{\chi} = F_{\chi}(\boldsymbol{\theta}, \boldsymbol{\epsilon_{c,B}}, \boldsymbol{\epsilon_{c,ADM}})$, and we can write

\begin{align}\label{eq: Bayes step 2}
    p(\boldsymbol{\theta}, \boldsymbol{\epsilon_{c,B}}, \nonumber \boldsymbol{F_{\chi}} |\mathbf{d}) \propto~& p(\boldsymbol{\theta}) p(\boldsymbol{\epsilon_{c, B}}|\boldsymbol{\theta}) p( \boldsymbol{F_{\chi}} |\boldsymbol{\theta}, \boldsymbol{\epsilon_c}) \\ & p\boldsymbol{(}\mathbf{d}|\mathbf{M}, \mathbf{R}\boldsymbol{)},
\end{align}
where $\boldsymbol{\epsilon_{c,B}}$ and $\boldsymbol{\epsilon_{c,ADM}}$ are the central energy densities of baryonic matter and ADM, respectively. 

Next, if we assume that each of the data sources are independent of one another and equate the likelihood to the PPM-derived mass-radius posteriors, Eq.~\ref{eq: Bayes step 2} becomes

\begin{align} \label{eq: Bayes step 3}
    p(\boldsymbol{\theta}, \boldsymbol{\epsilon_{c,B}}, \nonumber \boldsymbol{F_{\chi}} |\mathbf{d}) \propto~& p(\boldsymbol{\theta}) p(\boldsymbol{\epsilon_{c, B}}|\boldsymbol{\theta}) p( \boldsymbol{F_{\chi}} |\boldsymbol{\theta}, \boldsymbol{\epsilon_c}) \\ & \prod_{i} p(M_i, R_i ~|~d_{PPM,i}),
\end{align}
where $d_{PPM,i}$ is the mass-radius data derived from PPM and $i$ runs over the number of stars for which PPM delivers the mass and radius. Here, $p(M_i, R_i ~|~d_{PPM,i})$ is estimated by applying a Gaussian kernel density estimate to the \textit{NICER} PPM derived mass-radius data. The posterior distribution $p(\boldsymbol{\theta}, \boldsymbol{\epsilon_{c,B}}, \nonumber \boldsymbol{F_{\chi}} |\mathbf{d})$ is determined using the nested sampling software \texttt{MULTINEST} \cite{Feroz09, Feroz:2013hea,Buchner14}, where the prior is formed by drawing samples from the prior distribution on the EoS parameters to compute the corresponding $M$ and $R$, and the likelihood is calculated using the KDE of $p(M_i, R_i ~|~d_{PPM,i})$. Lastly, our framework neglects ADM halo configurations, which have been shown to modify the exterior spacetime of neutron stars and thus necessitate the modification of the PPM analysis pipeline used by \textit{NICER} \cite[see][]{Miao_2022,Shakeri2024,Shawqi:2024jmk}. To ensure that only ADM core configurations are computed in both the priors and resulting posteriors, we assign all halo configurations to have a zero likelihood.

Finally, to infer the properties of neutron stars with a MCDF core, and by extension the combined dark energy and baryonic EoS, the posterior distribution is modified in the following way

\begin{align}
    p(\boldsymbol{\theta}, \boldsymbol{\epsilon_c} |\mathbf{d}) &\propto p(\boldsymbol{\theta})  p(\boldsymbol{\epsilon_c}|\boldsymbol{\theta}) \prod_{i} p(M_i, R_i ~|~d_{PPM,i}).
\end{align}

The treatment of the central energy density differs between the ADM admixed neutron star model and the MCDF one is due to the difference in how we model their physical structures. In the ADM framework, the central energy density is defined by a vector of densities $\boldsymbol{\epsilon_c} = (\epsilon_b, \epsilon_{ADM})$, accounting for both the baryonic and ADM components. The MCDF model, however, assumes a dark energy core surrounded by a baryonic matter shell, which constrains the system to a single central energy density, $\epsilon_c$, for a given set of EoS parameters $\boldsymbol{\theta}$. Additionally, to make this work comparable to \citet{Pretel:2024vvt} and since the dark energy mass fraction inside neutron stars is unconstrained, the Bayesian framework with the MCDF EoS accounted for is identical to the one considering baryonic matter.

To ensure that no twin stars are formed, we impose that the central energy density satisfies $\boldsymbol{\epsilon_c} > \epsilon^+$. In addition, for all EoSs considered in this work, all priors and posteriors are computed with the constraint $M_i \geq 1 \, \Msun$. The constraint $M_i \geq 1 \, \Msun$ is chosen to be consistent with the X-ray Pulse Simulation and Inference \cite[X-PSI;][]{Riley23} PPM analysis pipeline, which was used to derive the \textit{NICER} mass-radius posteriors considered in this study (see Sec.~\ref{subsec: sources} for further details). Furthermore, the 1 $\Msun$ constraint is motivated  by the theoretical description of neutron stars early in their evolution \cite{Strobel_1999}. This minimum neutron star mass constraint is also in agreement with core-collapse supernova simulations \cite[see e.g.,][]{Radice2017, Suwa2018}.
    
\section{EoS priors \& source selection}\label{sec: priors and source selection}
In this section, we define the prior parameter ranges for the baryonic EoS, the bosonic and fermionic \citet{Nelson2019} ADM models, and the MCDF model. In addition, we discuss the astrophysical data sets used for our analysis.

\subsection{PP model priors}\label{subsec: PP priors}
As discussed in Sec.~\ref{subsec:baryonic eos}, we model the baryonic EoS using the PP model and  \citet{Keller2023} N$^3$LO $\chi$EFT calculations up to $1.5 n_0$. Here, the N$^3$LO $\chi$EFT band best-fit polytropic parameters discussed in Sec.~\ref{subsec:baryonic eos} are used as the prior ranges on the baryonic EoS for densities between [$0.572 n_0,\, 1.5n_0$].  To generate the PP model above $1.5n_0$, the prior ranges for the three polytropic indices are varied in the ranges of $\Gamma_1 \in [0,8]$, $\Gamma_2 \in [0,8]$, $\Gamma_3 \in [0.5,8]$, where the transition densities between the first to second polytrope ($n_1$) and the second to the third ($n_2$) are varied within $2.0 n_0 \leq n_1 \leq n_2 \leq 8.3 n_0$ \cite[see][for further details]{Rutherford:2024srk}. Furthermore, the first polytrope is used from $1.5 n_0$ to $n_1$, the second is used from $n_1$ to $n_2$, and the third from $n_2$ to the maximum central density allowed by causality. Lastly, we note that the PP model allows for first-order phase transitions with $\Gamma_2 = 0$ \cite[see][]{Hebeler2013,Greif19,Raaijmakers20}.

\subsection{Bosonic \& fermionic ADM priors}\label{subsec: ADM priors}
We now define the prior spaces for the fermionic and bosonic ADM parameters. For both ADM models, we define the priors on $m_\chi$, $g_\chi/m_\phi$, and $F_\chi$, all three of which fully characterize the \citet{Nelson2019} ADM model.

The prior ranges on the ADM self-repulsion strengths and particle masses are given by 

\begin{equation}
\frac{g_\chi}{m_\phi/\mathrm{MeV}} \in
\Bigg\{
    \begin{array}{lr}
        {[}10^{-2}, 10^{3}{]}, & \text{Bosonic ADM}\\
        {[}10^{-5},10^{3}{]}, & \text{Fermionic ADM}
    \end{array}
\end{equation}
\begin{equation}
m_\chi/\mathrm{MeV} \in
\Bigg\{
    \begin{array}{lr}
        {[}10^{-2}, 10^{8}{]}, & \text{Bosonic ADM}\\
        {[}10^{-2},10^{9}{]}, & \text{Fermionic ADM},
    \end{array}
\end{equation}
respectively. For bosonic ADM, the lower and upper bounds on $g_\chi/m_\phi$ were chosen to capture the values used by \citet{Nelson2019} and ensure the prior space has a finite size. The bounds on the bosonic $m_\chi$ prior were chosen such that no ADM particles could reach velocities larger than the escape velocity of a typical $1.4 \, \Msun$ neutron star and to evade black hole formation from bosonic ADM core collapse for the lower and upper limits, respectively \cite[see][and references therein]{Rutherford23}. 

For fermionic ADM, the lower bound on the $g_\chi/m_\phi$ prior is set by an approximation for $g_\chi/m_\phi = 0 \, \mathrm{MeV^{-1}}$, which is physically allowed due to the ADM fermionic degeneracy pressure, in order to satisfy our previous assumption of $g_B \ll g_\chi$ \cite[for details see][]{Rutherford:2024bli}. The upper bound on the fermionic $g_\chi/m_\phi$ is again set by the largest values used in \citet{Nelson2019} to keep the prior space bounded above. For the fermionic ADM $m_\chi$ priors, the bounds are set by the same physical limits used for the bosonic case, however, due to the fermionic ADM degeneracy pressure, fermionic ADM cores can support $m_\chi \leq 10^9$ MeV before gravitational collapse \cite{Gresham2018}. 

 The final ADM parameter for both of our ADM models is $F_\chi$, which, unlike the ADM particle mass, is not well constrained. Our previous works \cite{Rutherford23,Rutherford:2024bli} have discussed in detail the possible accumulation mechanisms, such as neutron bremsstrahlung \cite{Nelson2019}, neutron decay to ADM \cite{Ellis2018}, and accretion of ADM from the local density onto neutron stars \cite{Ivanytskyi2020}, in order to demonstrate the possible constraints on the ADM parameter space. However, two of the objectives of this work are to compare the inferences of ADM core to dark energy cores and to investigate how the ADM inferences change compared to our prior work on both models. Therefore, to illustrate the impact of a large ADM mass-fraction on neutron star inferences, while remaining consistent with the values and accumulation scenarios motivated in the current literature on dark matter admixed neutron stars \cite[see, e.g.,][]{Ellis2018, Miao_2022,Collier2022, Shawqi:2024jmk,Giangrandi2022,Thakur2024a, Shakeri2024,Mariani2024,Giangrandi2024,Avila2024, Cipriani:2025tga}, we take the upper bound on the prior, for both fermionic and bosonic ADM, to be
 \begin{equation}
     F_\chi \leq 5\%.
 \end{equation} 
\subsection{MCDF EoS priors}\label{subsec: CDF priors}
To generate the MCDF EoS, the prior intervals on the free parameters of the model must be defined. In particular, we use the works of \cite[][and references therein]{Pretel:2023nhf,Pretel:2024a,Pretel:2024tjw,Pretel:2024vvt} and any known physical constraints to define the priors on $A$, $\epsplus$, and $\alpha$. 

First, we consider the prior interval of the density jump parameter $\alpha$. The upper prior bound on $\alpha$ is directly determined by its definition, which is stated in Sec.~\ref{subsec:cdf eos}. We therefore take an upper prior bound of $\alpha \leq 1$. For the lower bound, we require $\alpha \neq 0$ because if $\alpha = 0$, the neutron star would be comprised entirely of dark energy, making the star undetectable by \textit{NICER}. We thus take the lower bound on the density jump parameter to be $\alpha \geq 0.1$ to ensure that the stars are visible by \textit{NICER} and to capture the values considered in \citet{Pretel:2023nhf,Pretel:2024tjw,Pretel:2024a}.

Next, we define the priors on $\epsplus$. Around the fiducial density of $1.5 n_0$, which corresponds to about $10^{14.60} \, \mathrm{g/cm^3}$, the \citet{Keller2023} N$^3$LO $\chi$EFT band imposes tight constraints on the baryonic EoS. To ensure that we consider neutron stars with baryonic shells extending to densities at least up to the maximum density in which we choose to trust $\chi$EFT, we impose that $\log_{10}(\epsplus \, \mathrm{cm^3/g}) \geq 14.60$. The upper bound is simply set by the maximum central energy density reached in a neutron star for a sampled EoS, which can reach values as high as $10^{16} \, \mathrm{g/cm^3}$. Combining the constraints imposed by $\chi$EFT and the maximum central density gives

\begin{equation}
    14.60 \leq \log_{10}(\epsplus \, \mathrm{cm^3/g}) \leq 16.
\end{equation}
However, as mentioned in Sec.~\ref{sec:inference framework}, we also impose the constraint $\epsplus < \epsilon_c$, which ensures that only neutron stars with dark energy cores are produced, thus the upper bound on the prior will shift depending on the sampled $\epsilon_c$.

The final free parameter of the MCDF EoS model is the dimensionless proportionality constant of the barotropic pressure term, $A$, which has been shown to result in appreciable changes in the mass-radius relation for values in the range $[0.2,0.7]$ \cite{Pretel:2024vvt}. For example, \citet{Pretel:2024vvt} showed that the maximum mass of a mass-radius curve can increase by about 20\% from $A=0.2$ to $A=0.3$. However, we find that neutron stars with masses greater than our imposed $1 \, \Msun$ constraint can be produced down to $A \approx 0.1$. For instance, an EoS defined by $A=0.1, \, \alpha = 0.9, \, \text{and} \, \log_{10}(\epsplus \, \mathrm{cm^3/g}) =14.62$, can produce neutron stars with masses up to $1.2 \, \Msun$. Therefore, we take the prior interval to be

\begin{equation}
    A\in [0.1,0.7].
\end{equation} 
Furthermore, we update the priors on $A$ to ensure the speed of sound within the MCDF core is causal. Thus, for each sampled EoS, we impose the constraint

\begin{equation}
    \frac{v_{s, \mathrm{MCDF}}^2}{c^2} = \frac{dP_{DE}}{d\epsilon_{DE}} = A + \frac{B}{\epsilon_{DE}^2} \leq 1,
\end{equation}
where $v_{s, \mathrm{MCDF}}$ is the speed of sound within the MCDF core.

\subsection{Astrophysical data sets}\label{subsec: sources}

\begin{figure}
    \centering
    \includegraphics[width=0.5\textwidth]{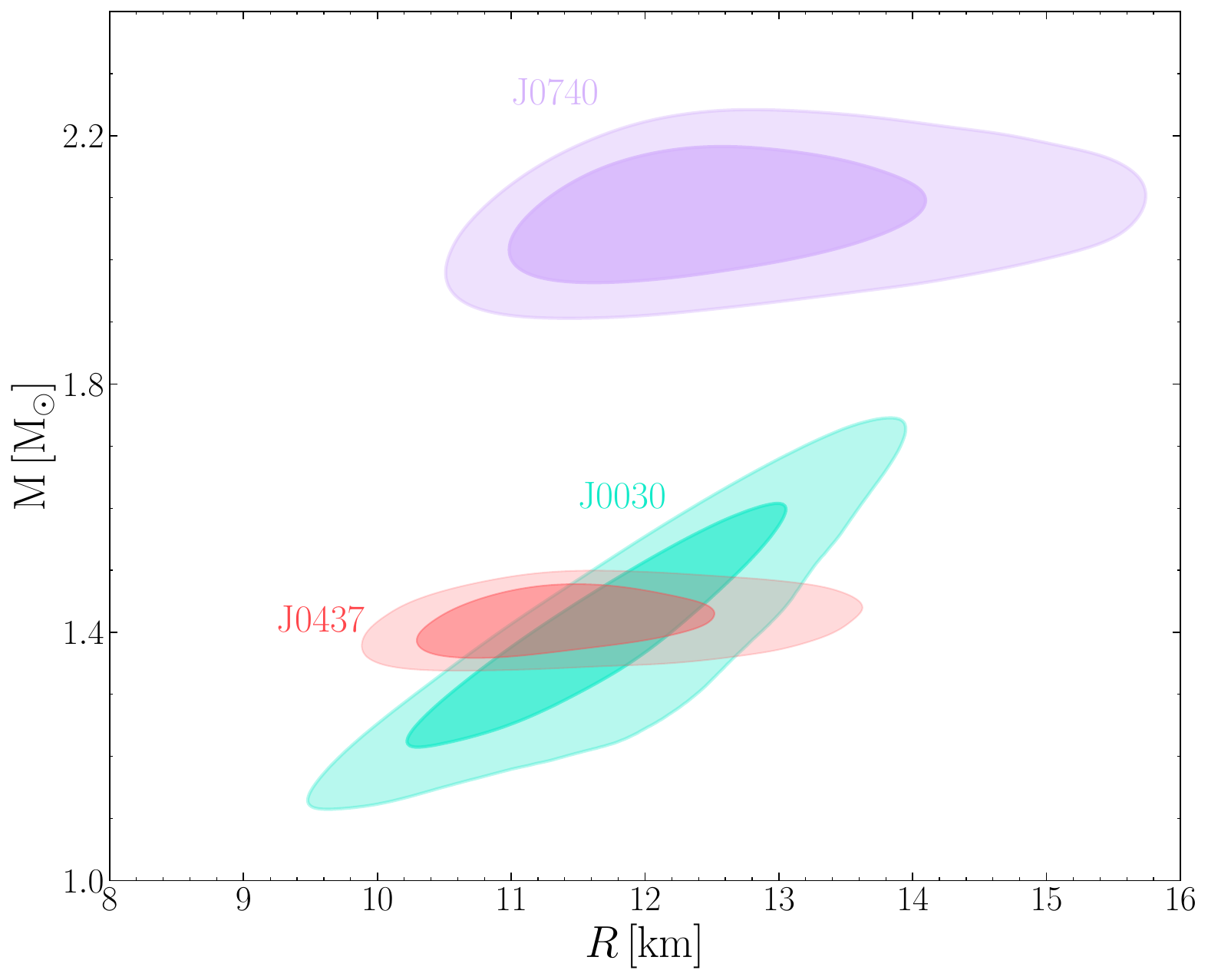}
    \caption{The 68\% (darker shaded region) and 95\% (lighter shaded region) confidence regions of the mass-radius posteriors of PSR~J0740 \cite{Salmi24}, PSR~J0030 \cite{Vinciguerra24}, and PSR~J0437\cite{Choudhury24}.}
    \label{fig:1}
\end{figure}

In our Bayesian framework, we incorporate the following \textit{NICER} mass-radius posteriors for three millisecond pulsars: PSR~J0740, PSR~J0030, and PSR~J0437. For PSR~J0740, we use the \citet{Salmi24} posteriors ($M = 2.07 \pm 0.07 \, \Msun$, $R = 12.49^{+1.28}_{-0.88}$ km at the 68\% confidence level), which benefit from an improved background treatment over earlier analyses. For PSR~J0030, we adopt the \texttt{ST+PDT} hot spot model from \citet{Vinciguerra24}, yielding $M = 1.40^{+0.13}_{-0.12} \, \Msun$ and $R = 11.71^{+0.88}_{-0.83}$ km\footnote{A recent paper by \citet{Kini26} updates this analysis, using a larger NICER data set together with XMM data and improved sampler settings. However, the results found in the new study are compatible with the results that we use from \citet{Vinciguerra24}.}. Lastly, for PSR~J0437, we include the posteriors reported in \citet{Choudhury24} ($M = 1.42 \pm 0.04 \, \Msun$, $R = 11.36^{+0.95}_{-0.63}$ km), which account for the 2017–2021 NICER data set. The joint $M-R$ posteriors on these three sources make up the astrophysical data considered in this work.

 \begin{figure*}
\centering
\includegraphics[width=\textwidth]{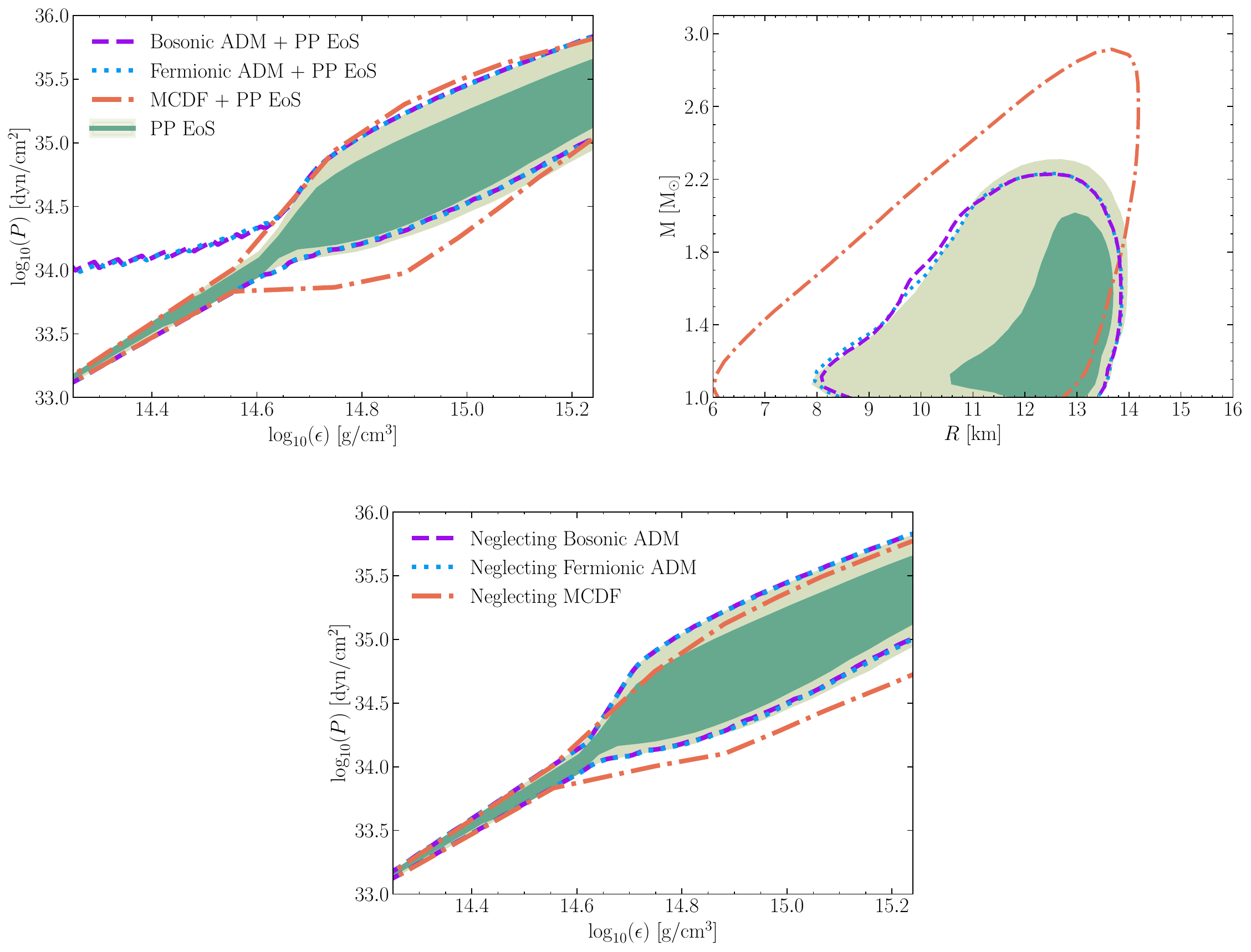}
\caption{Pressure-energy density prior distributions for the total combined dense matter EoS (top left), the resulting mass-radius prior distributions (top right), and the isolated baryonic EoS component (bottom. For the bottom panel, the contours are generated by extracting and plotting only the PP EoS pressure from the prior samples of each respective combined model. Across all panels, the light (dark) green bands represent the 68\% (95\%) CIs of the purely baryonic PP model. The purple dashed lines represent the 95\% prior of neutron stars with a possible bosonic ADM core, the light blue dotted lines represent the 95\% prior for a fermionic ADM core, and the orange dashed-dotted lines are the 95\% prior for stars with a MCDF core. Comparing the top left and bottom panels shows that, while the isolated baryonic pressures remain relatively unchanged when both bosonic and fermionic ADM are neglected, the total combined EoS pressure-energy density priors broaden significantly at lower densities when ADM is accounted for. On the other hand, including MCDF cores tends to push the isolated baryonic EoS towards lower pressures, whereas the total combined EoS pressure-energy density priors broaden towards lower pressures.  In the top right panel, we see the low-density pressure modification from the ADM cores favors lower maximum masses for the ADM admixed models, whereas the MCDF core priors significantly broaden toward higher masses at the 95\% level.}
\label{fig:2}
\end{figure*}

\section{Results: Impact of MCDF \& ADM Cores}\label{sec:results}
In this section, we investigate the inferred impacts of bosonic and fermionic ADM cores, and MCDF cores relative to the posterior inferences on the PP model. We first analyze the prior distributions for all four stellar models to understand the overall impact of considering the possibilities of ADM and MCDF cores inside neutron star interiors. Next, we analyze the posterior inferences for our four models to study the impact each has on the inferred the dense matter EoS and neutron star properties. We also investigate the posteriors of our models for the radii of $2 \, \Msun$ ($R_{2.0}$) and $1.4 \, \Msun$ ($R_{1.4}$) neutron stars. Lastly, we show the promise of constraining the MCDF EoS parameters, $A, \, \epsplus \, \text{and} \, \alpha$, using \textit{NICER} mass-radius results. While the  bosonic and fermionic ADM EoS models have been extensively investigated in our prior work \cite{Rutherford23,Rutherford:2024bli}, the analysis presented here incorporates additional observational constraints from PSR~J0437$-$4715, resulting in updated limits on the ADM parameter space. Thus, the full posterior distribution for the bosonic and fermionic ADM EoS parameters are provided for completeness in Appendix~\ref{sec:appendix_adm_posteriors}.

\subsection{Prior distributions}\label{subsec: prior dist}
Implementing the MCDF, bosonic and fermionic ADM, and PP equations of state generate the prior distributions shown in Fig.~\ref{fig:2}. In particular, Fig.~\ref{fig:2} shows, for each neutron star model, the prior distribution of the total combined EoS priors (top left panel), the resulting mass-radius priors (top right panel), and the isolated baryonic EoS (bottom panel). To isolate the effects of both ADM models and the MCDF model, respectively, on the PP EoS \textit{a priori}, the dark sector component was neglected and only the PP EoS pressure was computed for each EoS parameter vector sample from the combined ``Bosonic ADM $+$ PP EoS'', ``Fermionic ADM $+$ PP EoS'', and ``MCDF $+$ PP EoS'' priors. This results in the ``Neglecting Bosonic ADM'', ``Neglecting Fermionic ADM'', and ``Neglecting MCDF'' bands in the top left panels of Fig.~\ref{fig:2}, which can be understood as the prior distributions for the PP EoS when the dark sector components are neglected. When the bosonic and fermionic ADM components are neglected, the 95\% CI of the baryonic EoS remains mostly unchanged, with the lower end of the 95\% CI band shifting slightly to higher pressures. Moreover, both the `Neglecting Bosonic ADM' and `Neglecting Fermionic ADM' bands are identical to one another, indicating that neither fermionic nor bosonic ADM cores significantly affect the baryonic EoS. 

On the other hand, when MCDF cores are neglected, the resulting baryonic EoS shifts and broadens towards lower pressures, and as a result, softer EoSs. The softening of the isolated baryonic EoS results from our construction of neutron stars with an MCDF core. By modeling the stellar interior with a sharp interface separating the MCDF core from the surrounding baryonic shell, the baryonic EoS is required to match the MCDF EoS at this boundary. Since the interface occurs at a lower pressure than in the center of the star, this matching condition inherently selects softer EoS configurations in the prior. Stiffer baryonic EoS configurations require a higher pressure crossing with the MCDF EoS, which falls outside the physically viable MCDF parameter space for the selected prior ranges.

For the total combined pressure-energy density prior, Fig.~\ref{fig:2} shows that the presence of fermionic and bosonic ADM broadens the 95\% CI of the total neutron star EoS at densities below $\sim 10^{14.65} \, \mathrm{g/cm^3}$. The inclusion of ADM cores broadens the 95\% credible region of the total combined EoS towards higher pressures at densities below $\sim 10^{14.65} \, \mathrm{g/cm^3}$ because the ADM pressures are comparable to or greater than those of the PP model in this density regime. Beyond this density, the dominant pressure is provided by the PP model\footnote{We note that the total pressure is computed by summing the pressures from ADM and the baryonic EoS. More specifically, for the PP model, pressures are computed for each $\epsilon$ up to the maximum central baryonic density ($\epsilon_{\text{max} \, c,B}$) of the sampled PP EoS. For each sampled ADM EoS model, the ADM pressures are computed up to the maximum central ADM density corresponding to the sampled $F_\chi$ and $\epsilon_{\text{max} \, c,B}$. For the MCDF core neutron stars, however, the total pressure for the combined EoS is determined by computing pressures using the MCDF EoS for $\epsilon \geq \epsplus$ and the PP EoS for $\epsilon < \epsplus$.}. Quantitatively, at $\log_{10}(\epsilon \, \mathrm{cm^3/g}) = 14.4$, the upper limits of the 95\% CI of the total log-pressure priors for bosonic ADM, fermionic ADM, and the PP models are $33.93$, $33.94$, $33.59$, respectively. However, at $\log_{10}(\epsilon \, \mathrm{cm^3/g}) = 14.7$, we calculate that the 95\% CI upper limit of the total log-pressure priors are $33.35$, $33.39$, and $34.71$ for the bosonic ADM, fermionic ADM, and PP models, respectively. This shows that ADM cores contribute the dominant pressure at densities $\lesssim 10^{14.65} \, \mathrm{g/cm^3}$, but contribute negligibly to pressures above this density. From the observation that both ADM-admixed models broaden the 95\% CI of the total combined neutron star EoS at densities $\lesssim 10^{14.65} \, \mathrm{g/cm^3}$, but minimally deviate from the 95\% prior of the PP model at higher densities, we conclude that ADM cores widen the total dense matter EoS uncertainties at low density without impacting the EoS above these densities \textit{a priori}.

In mass-radius space, Fig.~\ref{fig:2} shows that the ADM admixed priors of both bosonic and fermionic ADM favor slightly lower maximum masses, but are otherwise very consistent with the PP model. This consistency with the PP model mass-radius priors is due to a strong degeneracy between the ADM EoS and PP model parameters. That is, a soft baryonic EoS can produce neutron stars identical to that of a relatively stiff baryonic EoS with an ADM core. As a result, we find that, under the current uncertainties of the baryonic EoS, ADM cores soften the overall neutron star EoS by favoring lower maximum masses. Finally, the prior distributions for bosonic and fermionic ADM are identical to each other in both panels of Fig.~\ref{fig:2}, suggesting that both ADM models affect the EOS in similar ways and, as a result, produce identical neutron stars.

\begin{figure*}
\centering
\includegraphics[width=\textwidth]{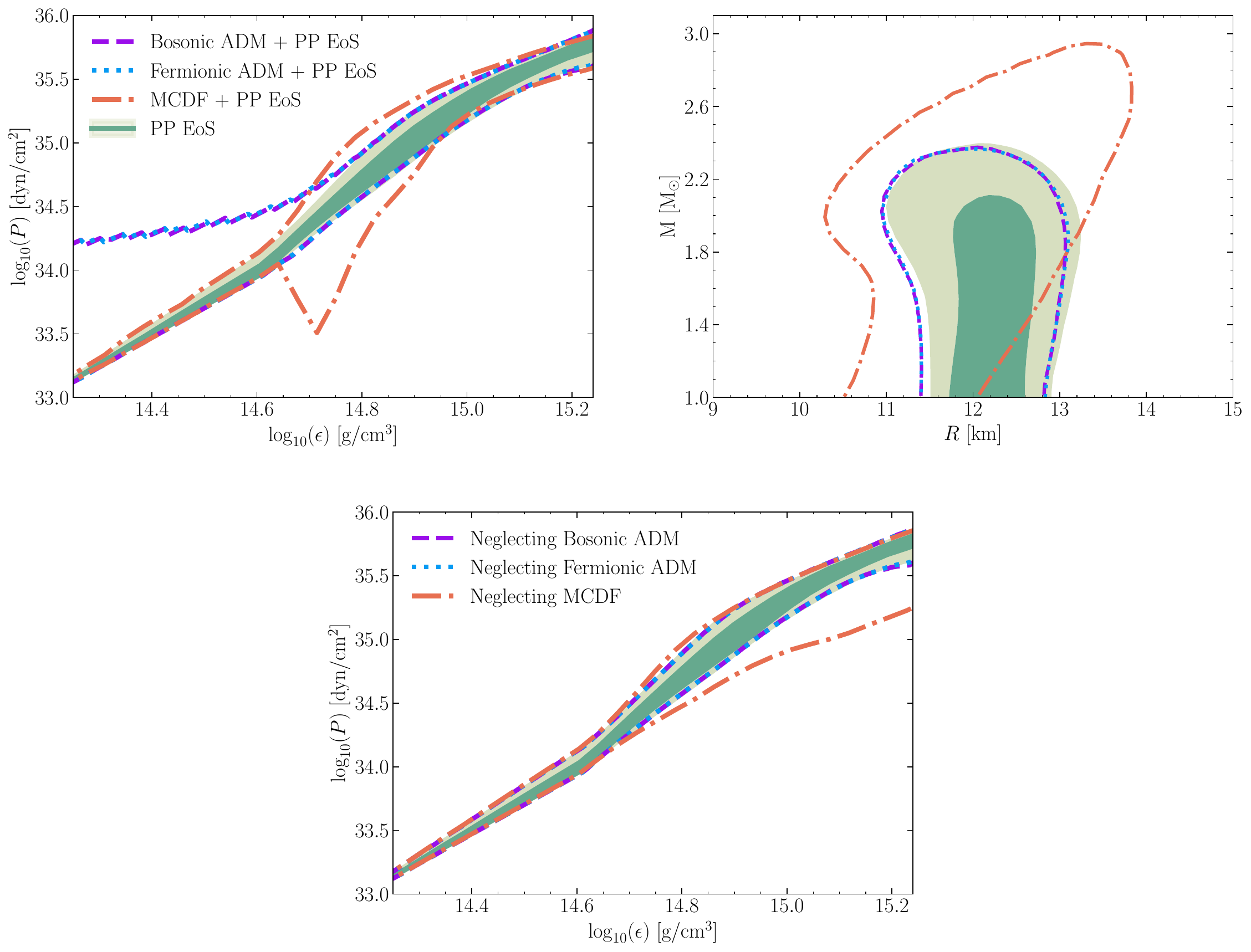}
\caption{Pressure-energy density posterior distributions for,the total combined dense matter EoS (top left), the total combined mass-radius (top right), and the isolated baryonic EoS (bottom) posterior distributions. The contour intervals and corresponding colors for each stellar model are the same as in Fig.~\ref{fig:2}. In all three panels, the MCDF core posteriors are significantly wider than those of the PP model and broaden the 95\% CI of the isolated baryonic EoS towards lower pressures at high density. Furthermore, the MCDF core neutron star model predicts the mean maximum mass to be $2.29 \, \Msun$. However, the total combined EoS posteriors of the ADM admixed models broaden at lower density and predict lower maximum masses for the total combined EoS (with a mean of $2.12 \, \Msun$), but do not change the posterior on the isolated baryonic EoS (which maintains a mean maximum mass of $2.16 \, \Msun$). The 95\% CI of the maximum mass of all 4 models is given in Table.~\ref{table 1}.}
\label{fig:3}
\end{figure*}

However, when considering neutron stars with MCDF cores and baryonic shells, Fig.~\ref{fig:2} shows that the total combined EoS prior, and by extension the corresponding mass-radius prior, significantly widens the 95\% CI towards lower pressure, compared to the PP and ADM admixed models. Including MCDF cores broadens the total EoS and mass-radius prior because, within our model, neutron stars are entirely described by the MCDF EoS within their cores, providing the sole source of pressure within this regime up to the transition density, $\epsplus$. Therefore, the combined neutron star pressure-energy density priors will be dominated by the uncertainties of $\epsplus$ in the MCDF model, which happens just above $10^{14.6} \, \mathrm{g/cm^3}$. Since the pressure-energy density priors significantly broaden and the mass-radius priors tend toward larger masses, we conclude that the possible presence of MCDF cores inside neutron stars can greatly alter the total dense matter EoS and neutron star observables.

\subsection{Inferred neutron star properties}\label{subsec: inferred of neutron star properties}

\begin{figure}
\centering
\includegraphics[width=.5\textwidth]{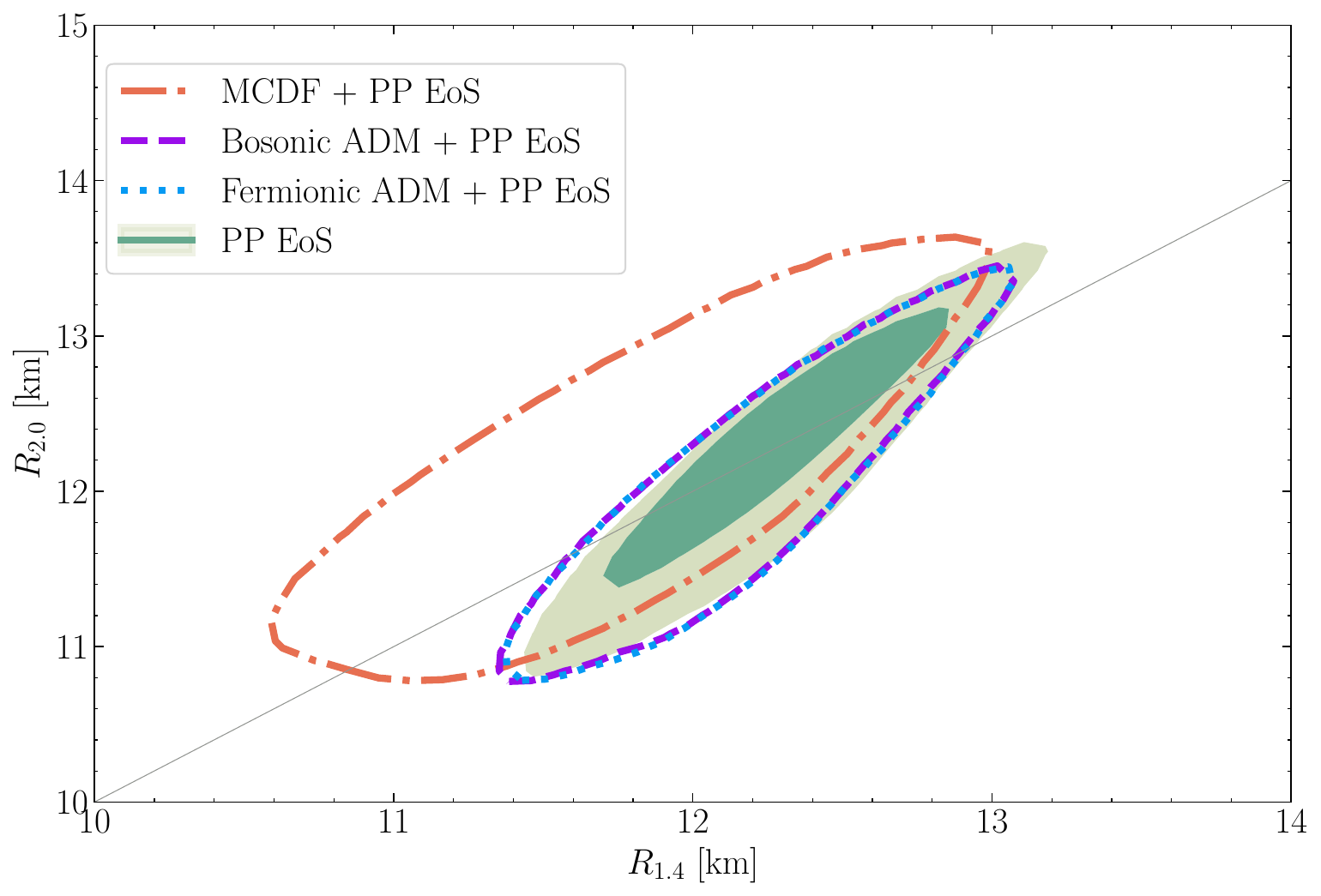}
\caption{Posterior correlation between the radius of a $2 \, \Msun$ ($R_{2.0}$) and $1.4 \, \Msun$ ($R_{1.4}$) neutron star for the PP model (light and dark green bands), bosonic ADM model (dashed purple), fermionic ADM model (blue dotted), and MCDF core model (dashed-dotted orange). The contour levels are the same as in Fig.~\ref{fig:2}. The solid gray lines represents the $R_{2.0} = R_{1.4}$ line. We find that the ADM admixed posteriors strongly overlap with the PP model, but the MCDF core posteriors widen along both the $R_{2.0}$ and $R_{1.4}$ axes, respectively.}
\label{fig:4}
\end{figure}

\begin{figure*}
\centering
\includegraphics[width=.75\textwidth]{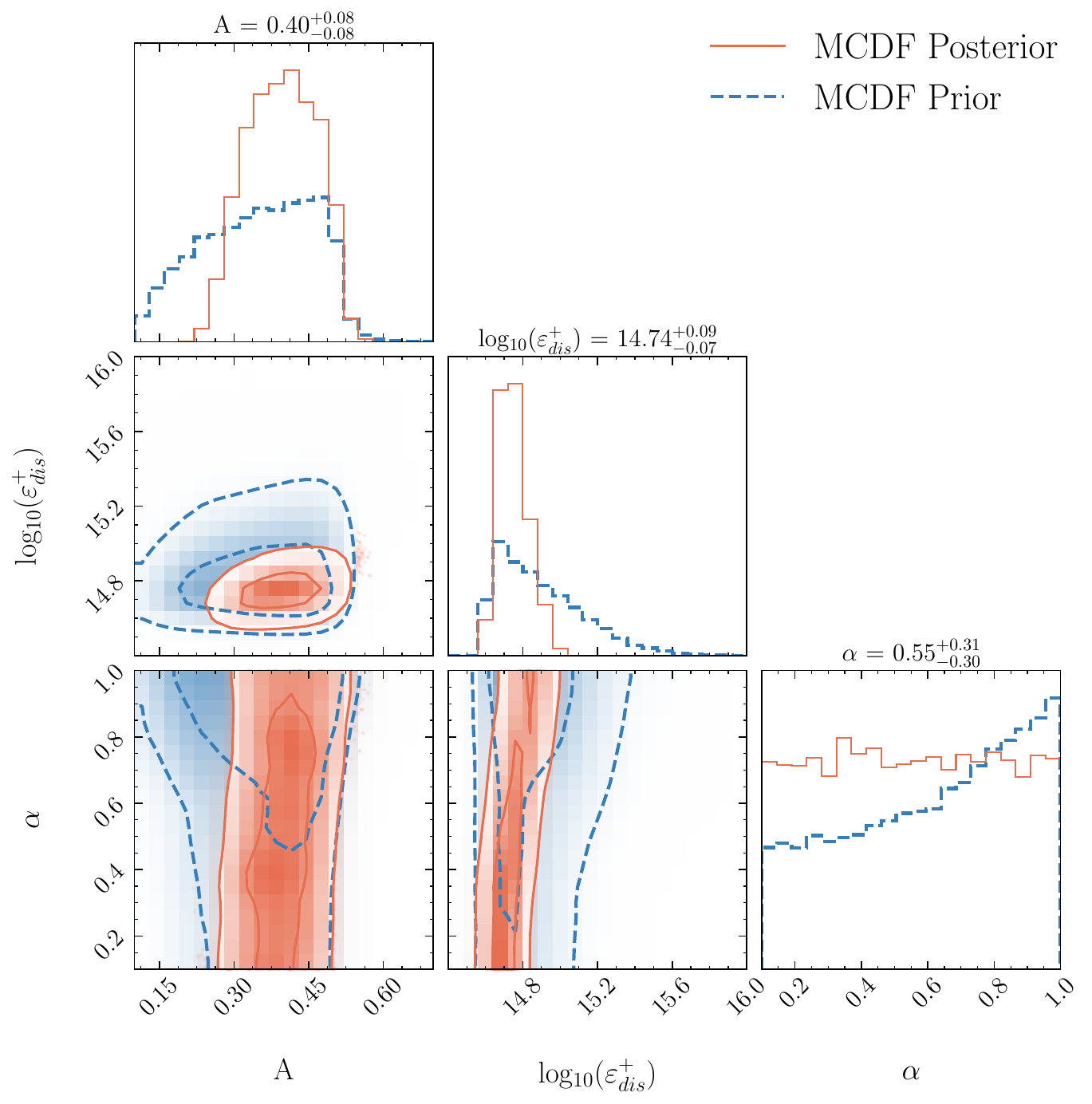}
\caption{Posterior (solid orange) and prior (dashed blue) corner plot of the MCDF EoS. The free parameters of the MCDF model, namely $A$, $\alpha$, $\epsplus$, are plotted against each other. The contour lines in the 2-D contour plots represent the 68\% and 95\% confidence regions. Along the diagonal, we show the 1-D histograms of each individual parameter, where the titles show the median value with the 1$\sigma$ uncertainties. Here, we see that the posterior distributions of $A$, $\epsplus$, their combinations, and those with $\alpha$, are narrowed compared to their corresponding priors.}
\label{fig:5}
\end{figure*}

Fig.~\ref{fig:3} shows the posterior distributions of the total combined EoS in the pressure-energy density space(top left panel), their corresponding mass-radius relation (top right panel), and the extracted baryonic EoS (bottom panel), for the PP model (dark and light green contours), stars admixed with fermionic and bosonic ADM cores (dotted blue and dashed purple lines, respectively), and stars with a MCDF core (orange dashed-dotted lines) in the top left and right panels, respectively. For the MCDF model, the top left panel of Fig.~\ref{fig:3} shows that the total pressure-energy density posteriors widen at the 95\% level for intermediate densities in the range between about $10^{14.6} \, \mathrm{g/cm^3}$ and $10^{15} \, \mathrm{g/cm^3}$, with the most substantial broadening occurring just above $10^{14.7} \, \mathrm{g/cm^3}$. Interestingly, the posteriors on $\epsplus$ peak near $10^{14.74} \, \mathrm{g/cm^3}$ and span a similar density range to the pressure-energy density posteriors (see Sec.~\ref{subsec: possible constraints on the CDF EoS} for details), suggesting that the neutron star EoS, and by extension the mass-radius relation, is particularly sensitive to the density of $\epsplus$ within our disjoint, non-interacting two-fluid model. A sensitivity to the transition density $\epsplus$ would broaden not only the pressure-energy density posteriors, but also the mass-radius posteriors. 

To illustrate this sensitivity,\footnote{We note that all illustrative calculations, namely the one mentioned here and the two described in Sec.~\ref{subsec: possible constraints on the CDF EoS}, consider the PP model described by following fixed parameters: $\Gamma_1 = 5$, $\Gamma_2 = 4$, $\Gamma_3 = 2.6$, $n_1 = 1.8 n_0$, $n_2 = 4 n_0$, $K_{\chi \text{EFT}} = 2.7$, and $\Gamma_{\chi \text{EFT}} = 2.624$.}, the maximum mass and radius of a neutron star mass-radius relation with a MCDF core described by $A = 0.3$, $\alpha = 0.6$, and $\log_{10}(\epsplus \, \mathrm{cm^3/g}) = 14.68$, is $2.18 \, \Msun$ and $12.02$ km, respectively. However, if we consider the same EoS but with $\log_{10}(\epsplus \, \mathrm{cm^3/g}) = 14.61$, the maximum mass and corresponding radius increase to $2.35 \, \Msun$ and $12.84$ km, respectively. Furthermore, because the $\log_{10}(\epsplus \, \mathrm{cm^3/g})$ posteriors favor densities just above the $\chi$EFT transition density of $14.6$, the resulting neutron stars will be composed of very large MCDF cores, which produce substantially higher masses than their purely baryonic counterparts.

In the mass-radius space, this broadening of the total combined pressure-energy posteriors translates into broader radii and much larger maximum masses at the 95\% CI when compared to both the PP model and the ADM admixed models. Outside of the density range of $\log_{10}(\epsilon \, \mathrm{cm^3/g}) \in [14.6,15.0]$, the MCDF model infers pressures similar to pressures of the PP model, thus showing the overall consistency between the two models. Since the MCDF model mass-radius and pressure-energy density posteriors strongly overlap with the PP model, but noticeably wider, we conclude that neutron stars with MCDF cores are equally as consistent with the \textit{NICER} data as the PP model, but predict substantially more massive neutron stars.

To better understand how the MCDF core influences the posterior distribution on the baryonic EoS, the pressure-energy density relation in which the MCDF core is neglected. As shown in the bottom panel of Fig.~\ref{fig:3}, isolating the baryonic EoS pressure shows the ``Neglecting MCDF'' 95\% CI extends preferentially towards lower pressures. The lower bound of ``Neglecting MCDF'' band encompasses a wider range of lower pressure values for the same reasons as observed in the prior: the baryonic EoS pressure must match the MCDF EoS pressure at $\epsplus$. Since the interface at $\epsplus$ occurs at a lower pressure than in the center of the star, the matching condition inherently selects softer EoSs. However, compared to the prior distribution which also includes a broad range of softer baryonic EOSs, the baryonic EoS component of the posterior distribution prefers stiffer EOSs to
support heavy-mass pulsars, namely that of PSR~J0740. Therefore, the ``Neglecting MCDF'' posterior's expansion into the low-pressure regime is not as low as that of the prior.

In contrast to the pronounced shifts resulting from the MCDF model, the ADM admixed core models exhibit more subtle deviations from the purely baryonic baseline of the PP EoS. Fig.~\ref{fig:3} shows that the 95\% level pressure-energy density posteriors broaden at densities below about $10^{14.75} \, \mathrm{g/cm^3}$, but otherwise predict pressures similar to the PP model. As mentioned in Sec.~\ref{subsec: prior dist}, this broadening results in lower maximum masses, but nearly identical radii to the PP model. Compared to their respective pressure-energy density prior, the ADM admixed model posteriors predict nearly identical pressures at the 95\% CI for densities $\lesssim 10^{14.75} \, \mathrm{g/cm^3}$. Furthermore, the bottom panel of Fig.~\ref{fig:3} shows that the both ADM models do not affect the posteriors of the baryonic EoS. From these observations, we find that, under the current uncertainties of the cold dense matter EoS, ADM admixed neutron stars are indistinguishable from baryonic neutron stars and that ADM core pressures cannot be constrained using the mass-radius measurements of PSR~J0740, PSR~J0030, and PSR~J0437.

Having evaluated the posteriors of all four models in the pressure-energy density and mass-radius spaces, respectively, we now examine their specific radius correlations to probe the high-density stiffness of the EoS. Fig.~\ref{fig:4} shows the joint posterior distribution on $R_{2.0}$ and $R_{1.4}$ for all four stellar models. Investigating  $R_{2.0}$ and $R_{1.4}$, and by extension $\Delta R = R_{2.0} - R_{1.4}$, can provide insight into the stiffness of the EoS at high densities \cite[see][]{Drischler2021}. More specifically, \citet{Drischler2021} showed that $\Delta R >0$($\Delta R < 0$), could be an indicator that the EoS stiffens (softens) at high densities. Here, Fig.~\ref{fig:4} and Table \ref{table 1} show that all mean values are consistent with $\Delta R$ being positive with the MCDF Model being the most positive\footnote{At first glance, the PP model $\Delta R$ posteriors appear to be in contrast with the values inferred in \citet{Rutherford:2024srk}. However, the posterior $\Delta R$ values were inferred using the additional posteriors of GW170817 \cite{gw170817}, and GW190425 \cite{gw190425}. Both of these jointly push the inferred $\Delta R$ to smaller values than the one quoted in this work.}. Compared to both the PP and ADM admixed models, the MCDF model widens the 95\% CI towards larger $R_{2.0}$ and smaller $R_{1.4}$, leading to $\Delta R$ being more positive. The inferred $\Delta R$ values of the MCDF core model are the most positive because the chosen MCDF core and baryonic shell contain a strong first-order phase transition, characterized by an abrupt transition density in the piecewise stellar EoS. Therefore, the resulting total EoS will have an abrupt change in overall stiffness when transitioning from the MCDF EoS to the baryonic EoS, which in turn changes the overall stiffness of the mass-radius relation and thus $\Delta R$. Although all of the stellar models favor mean $\Delta R$ values being positive, no conclusive statements on the stiffness of the neutron star EoS at high densities can be made due to the broad uncertainties.

\subsection{Constraints on the MCDF EoS parameter space}\label{subsec: possible constraints on the CDF EoS}

In Fig.~\ref{fig:5}, we show a corner plot of the posterior (solid orange) and prior (dashed blue) distributions on the MCDF EoS parameters, $A$, $\alpha$, and $\epsplus$.
Fig.~\ref{fig:5} shows that the MCDF posteriors of $A$ and $\epsplus$ are narrowed compared to their respective priors. In particular, the $\epsplus$ 1-D posterior is centered at the same density as the prior, but strongly favors densities in the range $\log_{10}(\epsplus \, \mathrm{cm^3/g}) \in [14.55,15]$. In addition, this range strongly overlaps with the broadening of the pressure-energy density posteriors of the MCDF model in Fig.~\ref{fig:3}, further demonstrating that the total neutron star EoS is highly sensitive to the value of $\epsplus$. For the $A$ parameter, the 1-D posterior is also centered near the median value of $A = 0.40$ and disfavors $A \lesssim 0.22$. To illustrate why these values of $\log_{10}(\epsplus)$ and $A$ are disfavored, we consider representative MCDF-cored neutron star models. A fiducial model with $A=0.3$, $\alpha = 0.6$, and $\log_{10}(\epsplus \, \mathrm{cm^3/g}) = 14.73$ produces $M_{\rm TOV} = 2.062 \, \Msun$, while reducing $A$ to $0.22$ or increasing $\log_{10}(\epsplus \, \mathrm{cm^3/g})$ to $15.03\,$ yields maximum masses of only $1.71 \, \Msun$ and $1.68 \, \Msun$, respectively. These reduced maximum masses are inconsistent with the observed mass measurement of PSR~J0740, thereby disfavoring these parameter regimes.

In contrast to the stringent constraints on $A$ and $\epsplus$, Fig.~\ref{fig:5} shows that the 1-D posterior of the density jump parameter is flat relative to the prior and thus does not disfavor any particular value. It is physically reasonable why no particular $\alpha$ value is inherently disfavored because the choice of density jump parameter changes the mass-radius relation by percentages smaller than than the uncertainties of the \textit{NICER} measurements of PSR~J0740, PSR~J0030, and PSR~J0437. For instance, consider two mass-radius curves, each described by $A = 0.3$ and $\log_{10}(\epsplus \, \mathrm{cm^3/g}) = 14.73$, but with $\alpha = 0.2$ and $\alpha = 0.9$, respectively. Both MCDF models generate similar masses, with the $\alpha = 0.2$ and $\alpha = 0.9$ models reaching maximum masses of $2.035 \, \Msun$ and $2.16 \, \Msun$, respectively. When comparing their respective radii for masses in the interval $[1 \, \Msun, 2 \, \Msun]$, we find that the average relative radial percent difference between these mass-radius curves is 13.07\%, while the \textit{NICER} measurements have uncertainties at the 10\% level. The observation that the posteriors of both $A$ and $\epsplus$ are substantially narrowed compared to their priors, while $\alpha$ is not, demonstrates that neutron star mass-radius measurements can tightly constrain the barotropic fluid constant and transition density of the MCDF EoS, even though the density jump parameter remains unconstrained. Moreover, under the current uncertainties of the baryonic EoS, these observations also show that \textit{NICER} can more tightly constrain dark energy cores than ADM cores.

\section{Summary \& conclusions}\label{sec:summary and conclusions}

In this work, we have explored the inferred constraints on the dense matter EoS using four different neutron star models, namely stars comprised of purely baryonic matter, an admixture of baryonic matter and fermionic ADM cores, an admixture of baryonic matter and bosonic ADM cores, and a dark energy core with a baryonic shell. Here we have modeled the baryonic EoS using the \citet{Keller2023} $\chi$EFT calculations up to $1.5 n_0$ and \citet{Hebeler2013} PP model at higher densities, both fermionic and bosonic ADM cores using the \citet{Nelson2019} ADM model, and the dark energy core using the MCDF model considered in \citet{Pretel:2024vvt}. To this end, we employed Bayesian parameter estimation on all four neutron star models using an adaptation of the \texttt{NEoST v2.2.0} inference code and considered the derived \textit{NICER} mass-radius measurements of PSR~J0740, PSR~J0030, and PSR~J0437 reported by \citet{Salmi24}, \citet{Vinciguerra24}, and \citet{Choudhury24}, respectively. By using four different neutron star models and the \textit{NICER} data of all three neutron stars, we inferred the effects of ADM cores and MCDF cores on neutron star observables and the promise of constraining the MCDF model parameter space. In Table~\ref{table 1}, we summarize the posterior results on $\mathrm{M}_{\rm TOV}$, $R_{2.0}$, $R_{1.4}$, and $\Delta R$.

 \begin{table*}
\caption{Posterior distributions on the maximum mass of a non-rotating neutron star $\mathrm{M}_{\rm TOV}$, the radius of a $2\, \Msun$ and a $1.4 \, \Msun$ neutron star, and their different $\Delta R = R_{2.0}-R_{1.4}$. The upper and lower values correspond to the 95\% CI of each quantity.}
\centering
\setlength{\extrarowheight}{4pt}
\begin{tabular}{|c|c|c|c|c|}
\hline
& PP model & MCDF Model& Fermionic ADM Model & Bosonic ADM Model\\
\hline
$\mathrm{M}_{\rm TOV}$ [$\Msun$] & $2.16^{+0.14}_{-0.15}$ & $2.29^{+0.45}_{-0.3}$ & $2.12^{+0.16}_{-0.14}$ & $2.12^{+0.16}_{-0.14}$ \\[1ex]
$R_{1.4}$ [km] & $12.26^{+0.65}_{-0.62}$ & $11.74^{+0.92}_{-0.82}$ & $12.16^{+0.64}_{-0.61}$ & $12.16^{+0.64}_{-0.62}$\\[1ex]
$R_{2.0}$ [km] & $12.27^{+0.92}_{-1.19}$ & $12.13^{+1.10}_{-1.06}$ & $12.09^{+0.93}_{-1.05}$ & $12.10^{+0.94}_{-1.06}$\\[1ex]
$\Delta R$ [km] & $0.02^{+0.33}_{-0.74}$ & $0.45^{+0.43}_{-0.84}$ & $0.07^{+0.37}_{-0.68}$ & $0.05^{+0.36}_{-0.67}$ \\[1ex]
\hline
\end{tabular}
\label{table 1}
\end{table*}

For neutron stars admixed with either bosonic or fermionic ADM cores, we find that the pressure-energy density 95\% CI priors of both models significantly broaden toward higher pressures at densities below $\approx 10^{14.65} \, \mathrm{g/cm^3}$ compared to the PP and MCDF models. Above this density, both ADM admixed pressure-energy density priors are nearly identical to the 95\% CI priors of the PP model. The total ADM pressure priors tend toward higher values at densities below $\approx 10^{14.65} \, \mathrm{g/cm^3}$ because the ADM core pressures in this regime exceed those of the PP model. At densities above $10^{14.65} \,\mathrm{g/cm^3}$, PP model pressures increase more rapidly than those of either ADM core model, causing the priors to converge toward the PP model pressures. 

Converting the bosonic and fermionic ADM priors to the mass-radius plane, we find that including the possible presence of ADM cores reduces the prior maximum mass by $\approx 0.1\, \Msun$ and strongly overlaps with the PP model priors at the 95\% CI. Accounting for ADM cores inside neutron star interiors reduces the maximum mass priors compared to the PP model because ADM cores decrease the masses and radii of neutron stars relative to identical stars with the same baryonic central energy density, thus shifting the ADM admixed mass-radius priors toward lower maximum masses. However, the entire ADM admixed mass-radius prior does not shift to lower radii due to the degeneracy between the baryonic and ADM EoSs, which implies that a stiff ADM EoS with a relatively soft baryonic EoS can generate similar mass-radius relations. These results show that the presence of ADM cores, whether bosonic or fermionic, relaxes the total neutron star EoS prior uncertainties at low density and reduces the prior maximum mass, but otherwise strongly overlaps with the PP model priors. Thus, we conclude that ADM cores soften the overall dense matter EoS and are fully consistent with stars composed of purely baryonic matter \textit{a priori}.

Moreover, since the fermionic and bosonic ADM priors strongly overlap across the entire pressure-energy density and mass-radius spaces, we conclude that both models generate identical admixed neutron star equations of state. This demonstrates that fermionic ADM admixed stars are indistinguishable from bosonic ADM admixed stars.

Upon applying our Bayesian framework and using the derived \textit{NICER} measurements to compute the posteriors, we find that the ADM admixed pressure-energy density posteriors broaden at densities below $10^{14.75} \, \mathrm{g/cm^3}$ and then converge to the posteriors of the PP model at high density. In addition, the broadening of the total pressure-energy posteriors is nearly identical to that of the priors. Since the ADM admixed EoS posteriors are identical to their priors for densities $\lesssim 10^{14.75} \, \mathrm{g/cm^3}$ and identical to the PP EoS posteriors at high densities, we find that the ADM admixed EoS cannot be constrained by current \textit{NICER} measurements, suggesting a possible need for measurements of lower mass neutron stars to probe this regime. We also find that the posterior on the isolated baryonic EoS is unaffected by the presence of ADM cores with $F_\chi \leq 5\%$, further supporting that neutron stars with ADM cores are equally as consistent with the \textit{NICER} data as stars composed of purely baryonic matter.

In the mass-radius space, the ADM admixed posteriors predict slightly lower maximum masses and strongly overlap with the PP model at the 95\% CI. In particular, we find the posterior 95\% CI on $\mathrm{M}_{\rm TOV}$ to be $2.16^{+0.14}_{-0.15} \, \Msun$ and $2.12^{+0.16}_{-0.14} \, \Msun$ for the PP model and ADM admixed models, respectively. Furthermore, we find that neutron stars with ADM admixed cores favor $\Delta R = R_{2.0} - R_{1.4}$ values nearly identical to those of the PP model at the 95\% CI. From these observations, we conclude that, similar to the findings of \cite{Rutherford:2024bli}, neutron stars with ADM cores with $F_\chi \leq 5\%$ are equally consistent with \textit{NICER} measurements as neutron stars without ADM. This implies that, under the current uncertainties of the baryonic EoS and for $F_\chi \leq 5\%$, ADM admixed neutron stars are indistinguishable from purely baryonic neutron stars.

The pressure-energy density and mass-radius priors of neutron stars with MCDF cores differ substantially from those with ADM cores and from the PP model. For the MCDF model priors, we find that the prior pressures relax at densities between $10^{14.6} \, \mathrm{g/cm^3}$ and $10^{15.2} \, \mathrm{g/cm^3}$ at the 95\% CI level. In the mass-radius space, we find that the presence of MCDF cores shifts the 95\% CI toward much higher masses for all radii. Within our disjoint, non-interacting two-fluid stellar model of a MCDF core and baryonic shell, the uncertainty of the neutron star EoS is dominated by the MCDF EoS parameter uncertainties, particularly that of $\epsplus$. Given that the prior on $\epsplus$ spans the range of the observed pressure broadening in the pressure-energy density prior, the total neutron star EoS uncertainty reflects the prior ranges of the MCDF EoS more so than that of the PP model, thus broadening the pressure-energy density 95\% credible regions. From the observation that the pressure-energy density and mass-radius priors greatly increase when MCDF cores are accounted for, we conclude that MCDF cores can significantly alter neutron star properties.

Converting the MCDF core model priors to their posteriors, we find that neutron stars with MCDF cores and the MCDF EoS parameters producing these cores are tightly constrained by \textit{NICER} observations. First, we find that the 1-D posteriors of $\epsplus$ and $A$ are strongly peaked and narrower compared to their priors. In particular, $\log_{10}(\epsplus \, \mathrm{cm^3/g}) \gtrsim 15.0$ and $A \lesssim 0.22$ are strongly disfavored. However, the transition density jump parameter $\alpha$ posterior is flat relative to the prior. Therefore, we conclude that \textit{NICER} can tightly constrain $\epsplus$ and $A$ of the MCDF EoS, while $\alpha$ cannot be constrained.

In the pressure-energy density plane, the posteriors of the MCDF model closely matches the PP model posteriors, but broadens in the log-energy density range of $[14.6,15.0]$. We also find that the 95\% CI of $\log_{10}(\epsplus \, \mathrm{cm^3 / g}) = 14.74^{+0.18}_{-0.14}$, which roughly spans the density range of this \mbox{broadening}, similar to the priors but much narrower.

Additionally, when isolating the baryonic EoS from the MCDF posterior, the ``Neglecting MCDF'' band extends toward lower pressures relative to the PP model at high density, although less severely than in the prior. This is because the baryonic EoS posterior favors stiffer configurations needed to support PSR~J0740, indicating that the matching condition at $\epsplus$ continues to bias the inferred baryonic EoS toward softer configurations even after folding in the \textit{NICER} mass-radius data.

Converting the pressure-energy density relation to mass-radius space, the posteriors favor substantially larger maximum masses with $\mathrm{M}_{\rm TOV} = 2.29^{+0.45}_{-0.30} \, \Msun$ at the 95\% CI level. We also find that the presence of MCDF cores shifts the mean posterior value of $\Delta R$ to $0.45$ km, albeit with large uncertainties, compared to the posteriors of the ADM admixed and PP models. The shift in the mean value of $\Delta R$, and the posterior radii more generally, is a direct consequence of the hybrid stellar model used to construct neutron stars with MCDF cores and baryonic shells. That is, because the disjoint, non-interacting two-fluid model contains a strong first-order phase transition from the dark energy core to the baryonic shell, the total neutron star mass-radius curve exhibits a rapid change in stiffness across the phase transition, thus shifting $\Delta R$ toward higher values. Since neutron stars with MCDF cores predict substantially larger $\mathrm{M}_{\rm TOV}$ and $\Delta R$, we conclude that stars with MCDF cores can be fully consistent with neutron star measurements and that \textit{NICER} can tightly constrain the corresponding EoS of these stars. Furthermore, the inferred maximum masses suggest that neutron stars with MCDF cores could offer an interesting explanation for some of the heaviest inferred compact object masses, such as that of PSR~J0952$-$0607 \cite{Romani:2022jhd} and the $2.6 \, \Msun$ compact object in GW190814 \cite{LIGOScientific:2020zkf}.

In conclusion, we have demonstrated an alternative approach to constraining possible models of dark energy using neutron star mass-radius measurements and Bayesian inference. We have also shown that ADM cores comprised of bosonic or fermionic ADM result in identical neutron stars and that neutron stars admixed with ADM cores are indistinguishable from purely baryonic stars for ADM mass-fractions less than 5\%, improving on the 1.7\% bound set by \citet{Rutherford:2024bli}. Furthermore, we find that \textit{NICER} observations can tightly constrain MCDF EoS parameter space, particularly the transition density $\epsplus$ and barotropic fluid constant $B$, while MCDF cores predict substantially higher maximum masses that may explain some of the heaviest observed compact objects. Additionally, we find that all four neutron star models have positive mean values of $\Delta R = R_{2.0} - R_{1.4}$, with the MCDF model showing the most pronounced shift toward stiffening at high density. However, given the broad uncertainties on $\Delta R$, these results do not provide a conclusive statement on the high-density stiffness of the neutron star EoS for these four models.

Finally, several avenues of future work remain open. First, the ADM priors derived in this work assume ADM exists only in the stellar core. However, ADM may also form halo configurations that extend through and beyond the baryonic surface of a neutron star. In this scenario, the ADM halo introduces additional gravitational lensing effects, modifying the X-ray flux observed by \textit{NICER} and thus altering the inferred mass-radius posteriors from the PPM analysis pipeline. It would therefore be valuable to assess how the EoS-informed pulse profile modeling pipeline of \citet{Hoogkamer:2025ype}, which uses normalizing flows to incorporate the mass-radius priors of the EoS into the X-PSI PPM framework, is modified when ADM priors that fully account for both core and halo configurations are considered. Second, all four neutron star models considered in this work were computed in the non-rotating limit. Since PSR~J0740, PSR~J0030, and PSR~J0437 are millisecond pulsars, re-performing this Bayesian analysis within a rapidly rotating (two-fluid) framework represents another interesting direction because \citet{Konstantinou2024} showed that the universal relations governing how mass and radius changes due to rotation remain largely unaffected by the presence of DM cores for mass fractions up to 5\%. This suggests that these relations could be employed to extend the present inference to rotating configurations. However, when ADM halos are accounted for, the ADM component can be assumed to be torque-free and differentially rotating, while the baryonic matter rotates rigidly \citep{Shawqi:2025cca}, introducing structural complexity beyond what the mass-radius universal relations might be able to capture. Since rotation systematically shifts mass-radius curves to higher masses and larger radii, incorporating rotation into our framework would make our analysis more physically consistent with the PPM data analysis pipeline and could modify the constraints on both the MCDF and ADM parameter spaces. Altogether, this analysis provides a foundation for future studies, adding to the growing consensus that Bayesian inference techniques are essential for disentangling the interactions between the dark sector and neutron star observables.

\begin{figure*}
\centering
\begin{subfigure}{0.50\textwidth}
   \includegraphics[width=\textwidth]{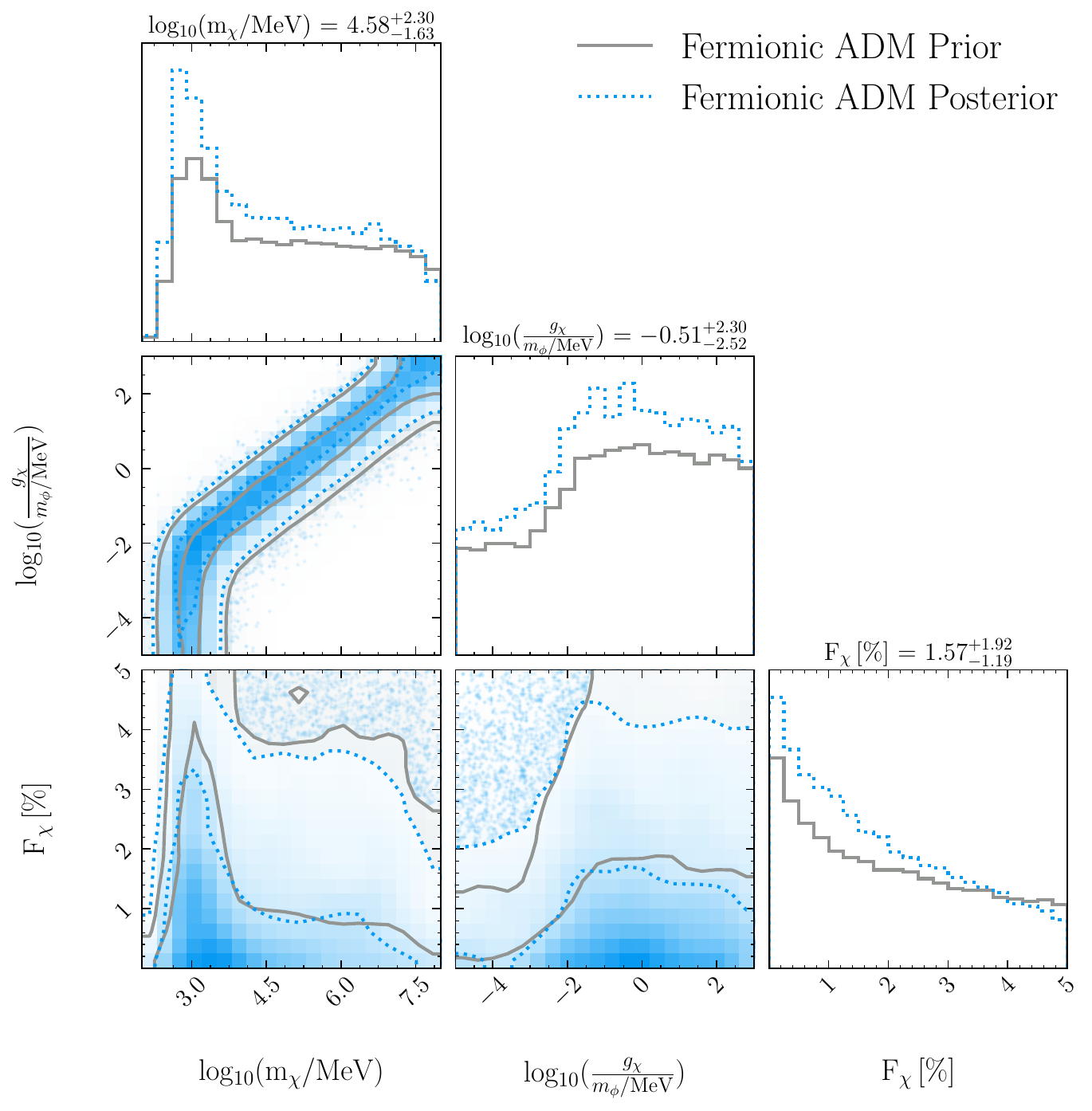}
   \label{appendix:fig 1a} 
\end{subfigure}%
\begin{subfigure}{0.50\textwidth}
   \includegraphics[width=\textwidth]{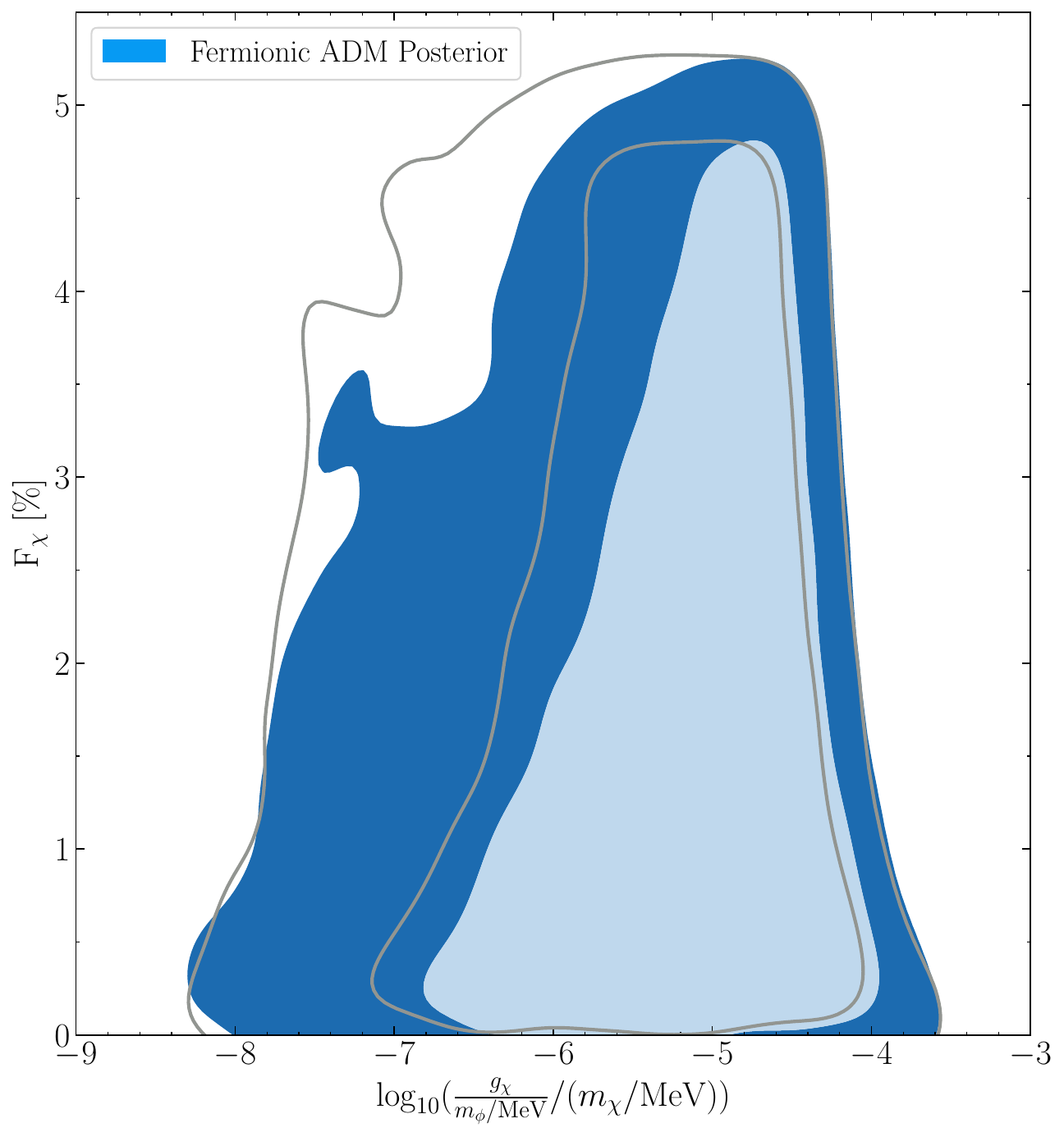}
   \label{appendix:fig 1b}
\end{subfigure}
\caption{Left panel: Posterior (dotted blue lines) and prior (solid grey lines) distributions of the fermionic ADM EoS
parameters. The contour levels are same as in Fig.~\ref{fig:5}; Right panel: Probability density contour plot of the fermionic ADM posteriors in the $F_\chi$-$\log_{10} \left(\frac{g_\chi}{(m_\phi/\mathrm{MeV})}/(m_\chi/\mathrm{MeV})\right)$ plane. Note, the contours represent the 1$\sigma$ (light blue) and 2$\sigma$ (dark blue) levels for both the prior and posterior. In the right panel, the posterior ratio of $g_\chi/m_\chi$ and $m_\chi$ grows more narrow for increasing $F_\chi$ at both the 1$\sigma$ and 2$\sigma$ contour levels, indicating tight constraints can be imposed on the ratio if neutron stars have large $F_\chi$ near 5\%. However, the left panel shows that the individual fermionic ADM posteriors are nearly identical to their respective priors, showing that they cannot be constrained.}
\label{appendix:fig 1}
\end{figure*}

\section*{Acknowledgments}
We acknowledge Ann Nelson for her foundational work on dark matter admixed neutron stars that made this work possible. A.L.W.~acknowledges support from NWO ENW-XL grant OCENW.XL21.XL21.038 {\it Probing the phase diagram of Quantum Chromodynamics}. C.P.W. acknowledges all the administrative and facilities staff at the University of New Hampshire, especially Katie Makem-Boucher and Michelle Mancini. The contributions of C.P.W.~and N.R.~were supported by NASA grant No.80NSSC22K0092. The work of C.P.W.~and N.R.~were also supported in part by grant NSF PHY-2309135 to the Kavli Institute for Theoretical Physics (KITP). N.R.~also acknowledges support from the Foundational Questions Institute (FQxI) and the University of New Hampshire Dissertation Fellowship. We acknowledge extensive use of NASA’s Astrophysics Data System (ADS) Bibliographic Services, the INSPIRE-HEP database, and ArXiv.

\textit{Software}: Python/C language~\cite{Oliphant2007}, GNU~Scientific~Library~\cite[GSL;][]{Gough2009}, NumPy~\cite{vanderWalt2011}, Cython~\cite[][]{Behnel2011}, SciPy~\cite{Virtanen2020}, MPI for Python~\cite{Dalcin2008}, Matplotlib~\cite{Hunter2007}, Jupyter~\cite{Kluyver2016}, MultiNest~\cite{Feroz:2013hea}, \textsc{PyMultiNest}~\cite{Buchner14}, kalepy~\cite{Kelley2021}, corner~\cite{ForemanMackey2016}, seaborn~\cite{Waskom2021}, NEoST~\cite{Raaijmakers:2025hbz}.

\begin{figure*}
\centering
\begin{subfigure}{0.50\textwidth}
   \includegraphics[width=\textwidth]{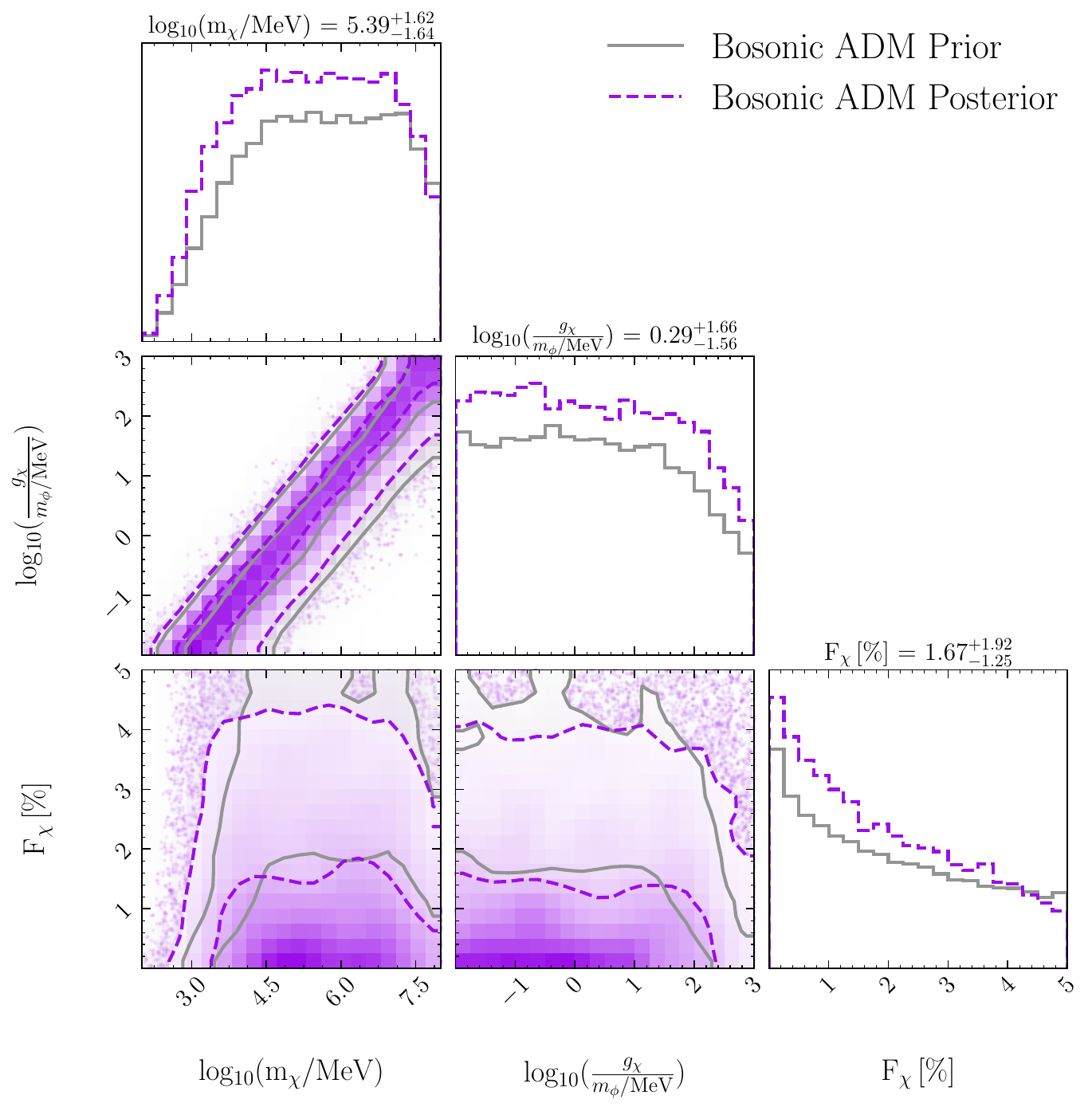}
   \label{appendix:fig 2a} 
\end{subfigure}%
\begin{subfigure}{0.50\textwidth}
   \includegraphics[width=\textwidth]{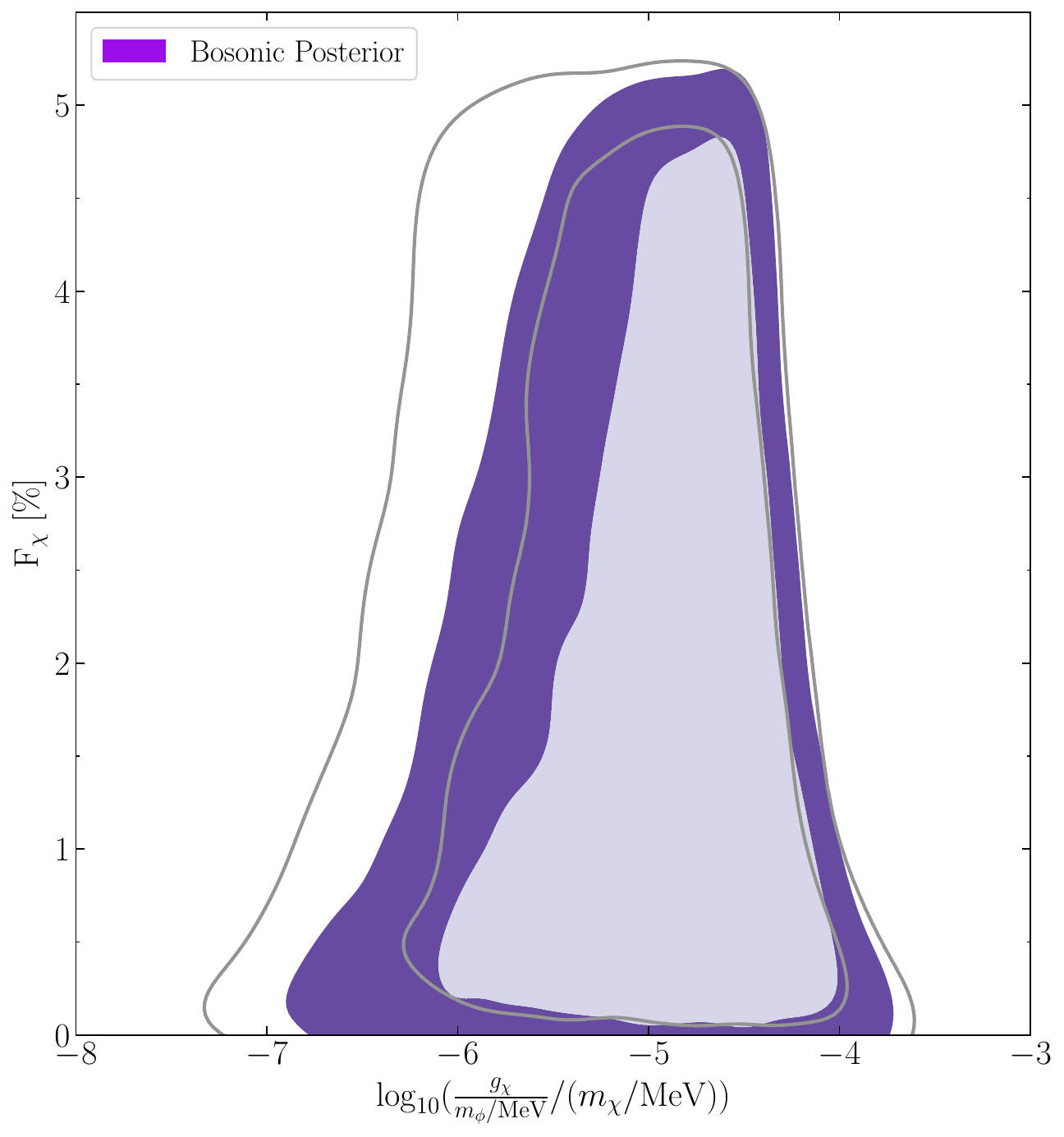}
   \label{appendix:fig 2b}
\end{subfigure}
\caption{Same as Fig.~\ref{appendix:fig 1}, but for bosonic ADM. The corner plot shows that the ratio of $\log_{(g_\chi/m_\chi)}$ and $\log_{10}(m_\chi)$ can be slightly constrained, but individually all bosonic ADM parameters are unconstrained relative to their respective priors. While in the right panel, $\log_{10} \left(\frac{g_\chi}{(m_\phi/\mathrm{MeV})}/(m_\chi/\mathrm{MeV})\right)$ can be more tightly constrained for increasing $F_\chi$, at both 1$\sigma$ and 2$\sigma$ contour levels. We also find that the posterior ratio of $g_\chi/m_\chi$ and $m_\chi$ for bosonic ADM can be more tightly constrained for all $F_\chi$ than that of fermionic ADM.}
\label{appendix:fig 2}
\end{figure*}

\appendix
\section{Bosonic and Fermionic ADM Posteriors}\label{sec:appendix_adm_posteriors}

Fig.~\ref{appendix:fig 1} and \ref{appendix:fig 2} show the posterior distributions for the fermionic and bosonic ADM EoS parameters, respectively. The corner plots of Fig.~\ref{appendix:fig 1} and \ref{appendix:fig 2} (left panels) show that the posterior logarithmic ratio of $g_\chi/m_\phi$ to $m_\chi$ of both ADM models is only marginally more narrow than its priors. Individually, the fermionic and bosonic ADM EoS parameter posteriors are identical to their respective prior, indicating that the ADM EoS parameters cannot be constrained individually using the current \textit{NICER} data.

As illustrated in Figs.~\ref{appendix:fig 1} and \ref{appendix:fig 2}, the $F_\chi$-$\log_{10} \left(\frac{g_\chi}{(m_\phi/\mathrm{MeV})}/(m_\chi/\mathrm{MeV})\right)$ space reveals that higher values of $F_\chi$ lead to a narrower lower bound for the $(g_\chi/m_\phi)/m_\chi$ posteriors (at both 68\% and 95\% confidence levels). The upper bounds, however, are unaffected by the data and continue to match the priors for both ADM models. As \citet{Rutherford:2024bli} found, the lower bound on the ratio of $g_\chi/m_\phi$ to $m_\chi$ is more constrained for increasing $F_\chi$ because small ratios produce compact ADM cores with high ADM central densities, thus reducing the maximum neutron star rapidly as $F_\chi$ increases. On the other hand, for a given $F_\chi$, a larger ratio of $g_\chi/m_\phi$ to $m_\chi$ (e.g., $\log_{10} \left(\frac{g_\chi}{(m_\phi/\mathrm{MeV})}/(m_\chi/\mathrm{MeV})\right) \approx -4$) near the upper limits of the prior produces more diffuse ADM cores with low ADM central densities compared to that of baryonic matter, which affects the neutron star mass-radius less significantly and therefore can fit \textit{NICER} data more easily. 

The right panels of Figs.~\ref{appendix:fig 1} and \ref{appendix:fig 2} also show that the bosonic ADM posterior ratio of $g_\chi/m_\chi$ to $m_\chi$ is more tightly constrained for all $F_\chi$ than the posterior ratio for fermionic ADM. Bosonic ADM is more sensitive to the ratio of self-repulsion to ADM particle mass for increasing $F_\chi$ than fermionic ADM. This is due to the fact that the bosonic ADM pressure is entirely determined by $g_\chi/m_\phi$, where as the fermionic ADM pressure has an additional term from the fermi degeneracy pressure. Therefore, as $F_\chi$ increases for a given ratio, the bosonic ADM core will become too compact to satisfy the \textit{NICER} data for a given increase in $F_\chi$ than that of fermionic ADM. More quantitatively, we find that the bosonic ADM posterior is constrained to be 
\begin{align}
\log_{10} & \left( \frac{g_\chi}{(m_\phi/\mathrm{MeV})(m_\chi/\mathrm{MeV})} \right) \nonumber \\ &\geq -4.94^{+0.82}_{-1.42}(-4.94^{+0.47}_{-0.71})
\end{align}
at the 95\% (68\%) confidence level. However, the fermionic ADM posterior is constrained to 
\begin{align}
\log_{10} & \left( \frac{g_\chi}{(m_\phi/\mathrm{MeV})(m_\chi/\mathrm{MeV})} \right) \nonumber \\ &\geq -5.26^{+1.10}_{-2.40}(-5.26^{+0.69}_{-1.08})
\end{align}
at the 95\% (68\%) confidence level. From the observation that the posterior ratio of $g_\chi/m_\chi$ to $m_\chi$ narrows for increasing $F_\chi$ for both bosonic and fermionic ADM, we conclude that \textit{NICER} mass-radius measurements can constrain the ratio of $g_\chi/m_\chi$ to $m_\chi$ with bosonic ADM exhibiting tighter constraints.

\newpage
\bibliography{eos}
\end{document}